\documentclass[twocolappendix, twocolumn, tighten, aps, revtex4, amsmath,amssymb, floatfix]{openjournal}

\usepackage{layouts}
\usepackage{natbib}
\setcitestyle{authoryear,open={(},close={)}}
\usepackage[dvipsnames]{xcolor}
\definecolor{mycitecolor}{HTML}{0071BC}
\definecolor{mybibcolor}{HTML}{15998E}
\usepackage[colorlinks=true,linkcolor=black,citecolor=mycitecolor, urlcolor=mybibcolor]{hyperref}%
\def\equationautorefname~#1\null{Equation (#1)\null}

\usepackage{graphicx}
\usepackage{dcolumn}
\usepackage{bm}
\usepackage{amsthm}
\usepackage{algorithmic}
\usepackage{centernot}
\usepackage[normalem]{ulem}
\usepackage{svg}
\usepackage{xprintlen}
\usepackage{adjustbox}
\usepackage{booktabs}

\usepackage{ragged2e}

\usepackage{afterpage}
\newcommand\myemptypage{
    \null
    \thispagestyle{empty}
    \addtocounter{page}{-1}
    \newpage
    }
    
\providecommand{\keywords}[1]
{
  \small	
  \textbf{\textit{Keywords---}} #1
}

\newcommand{\dat}{\textbf{d}}
\newcommand{\parvec}{\boldsymbol{\theta}}
\newcommand{\xvec}{\textbf{{x}}}
\newcommand{\muvec}{\boldsymbol{\mu}}

\newcommand{\myglobal}{\textbf{u}}
\newcommand{\edge}{\textbf{e}_k}

\newcommand{\node}{{\textbf{v}}}
\newcommand{\nodei}{{\textbf{v}}_i}

\usepackage{amsmath}
\DeclareMathOperator{\arcsinh}{arcsinh}
\DeclareMathOperator{\tr}{tr}

\usepackage[normalem]{ulem}

\begin{document}

\title{The Cosmic Graph: Optimal Information Extraction from Large-Scale Structure using Catalogues}

\author{T. Lucas Makinen}
\email{l.makinen21@imperial.ac.uk}

\affiliation{%
    Imperial Centre for Inference and Cosmology (ICIC) \& Astrophysics Group,
    Imperial College London, Blackett Laboratory, Prince Consort Road, London SW7 2AZ,
    United Kingdom
}%

\affiliation{Harvard \& Smithsonian Center for Astrophysics,
Observatory Building E, 60 Garden St, Cambridge, MA 02138, United States}

\author{Tom Charnock}
\affiliation{%
    Freelance consultant in statistical modelling
}%

\author{Pablo Lemos}

\affiliation{%
Department of Physics and Astronomy, University of Sussex, Brighton, BN1 9QH, UK
}%
\affiliation{University College London, Gower St, London, UK}


\author{Natalia Porqueres \& Alan Heavens}%
\affiliation{%
Imperial Centre for Inference and Cosmology (ICIC) \& Astrophysics Group,
Imperial College London, Blackett Laboratory, Prince Consort Road, London SW7 2AZ,
United Kingdom
}

\author{Benjamin D. Wandelt}
\affiliation{%
 Sorbonne Université, CNRS, UMR 7095, Institut d’Astrophysique de Paris, 98 bis boulevard Arago, 75014 Paris, France
}%
\affiliation{%
Center for Computational Astrophysics, Flatiron Institute, 162 5th Avenue, New York, NY 10010, USA
}%


\begin{abstract}
We present an implicit likelihood approach to quantifying cosmological information over discrete catalogue data, assembled as graphs. To do so, we explore cosmological { parameter constraints} using mock dark matter halo catalogues. We employ Information Maximising Neural Networks (IMNNs)
to quantify Fisher information extraction as a function of graph representation. We a) demonstrate the high sensitivity of modular graph structure to the underlying cosmology in the noise-free limit, b) show that {graph neural network summaries} automatically combine mass and clustering information through comparisons to traditional statistics, c) demonstrate that networks can still extract information when catalogues are subject to noisy survey cuts, and d) illustrate how nonlinear IMNN summaries can be used as asymptotically optimal compressed statistics for Bayesian simulation-based inference. We reduce the area of joint $\Omega_m, \sigma_8$ parameter constraints with small ($\sim$100 object) halo catalogues by a factor of 42 over the two-point correlation function, and demonstrate that the networks automatically combine mass and clustering information. This work utilizes a new IMNN implementation over graph data in Jax, which can take advantage of either numerical or auto-differentiability. We also show that graph IMNNs successfully compress simulations away from the fiducial model at which the network is fitted, indicating a promising alternative to $n$-point statistics in catalogue {simulation-}based analyses.

\end{abstract}


\maketitle

\section{\label{sec:intro}Introduction}
Modern cosmological analyses typically focus on obtaining theory and parameter constraints from compressed summary statistics obtained from field data such as the Cosmic Microwave Background or weak lensing mass-maps \citep{ Tegmark_1997,alsing2018_general, Jeffrey_2020LFI_field}. Recently, field-level analyses like \cite{natalia2021, leclercq2021accuracy}, although computationally expensive, have made it possible to sample the full field at the pixel level to ensure all survey information is accounted for in posterior construction for cosmological parameters.

However, the data collected by telescopes are often instantly compressed into discrete catalogues of sources, like galaxies and their underlying dark mater halos, or cosmic voids \citep{2012ApJ...761...44S,kreisch_voids}. The typical approach taken to analyse galaxy cluster data is to ``paint'' identified sources onto a grid and perform luminosity peak counts in high-density regions as a tracer for underlying dark matter. Analyses of these catalogues usually focus on 2-point information, either in real or Fourier space. However, these statistics are only sufficient when the underlying field is Gaussian, which is not the case for late-time cosmic web structures. Finding a statistic with which to capture {more} of this information is an active area of research. Existing methods include the three-point correlation function (the bispectrum in Fourier space, e.g. \cite{philcox_bispectra2022}), Minkowski functionals \citep{petri_minkowski}, the 1D probability distribution function \citep{uhlemann_1dprob}, marked power spectra \citep{marked_pk_quijote}, minimum spanning trees \citep{barrow_mst_1985,naidoo_mst_2020, cosmo_from_graphsNaidoo_2022}, and field-level sampling \citep{2013MNRAS.432..894J,2019A&A...621A..69R,natalia2021, leclercq2021accuracy, florent_borg, Jasche_2015}. However, truncating analyses to power or bispectra almost certainly discards information, especially for highly non-Gaussian fields, while field-level methods quickly become computationally expensive with increasing survey volume.
Likewise, void cosmology constructs correlation functions from void positions and redshifts \citep{hamaus_voids2015} to capture under-dense regions in structure formation. This sort of analysis usually discards morphological features of voids, such as void ellipticity, resulting in a loss of information that could be relevant to the underlying cosmological model \citep{2010PhRvD..82b3002B,2010MNRAS.403.1392L,findvoids_topology2019}.

Graphs provide a natural way to describe the nonlinear aspects of large-scale structure (LSS). Dark matter halos and their galaxy clusters can be attributed to nodes (vertices), while filaments are traced by smaller halos and edges connecting neighbouring edges. In this representation, clustering under gravity can be translated into higher connectivity or number of edges. Higher order $n$-point functions can be computed efficiently for clusters, while avoiding the cost of computing extraneous connections across voids. Graph representation of LSS promises a more modular approach to information quantification, and compliments the existing body of literature. Minimum spanning trees (MSTs) have been used in cosmological analyses since \cite{barrow_mst_1985}, and subsequent studies have investigated using binned halo graph features from simulations as cosmological probes \citep{bhavsar_1988mst, vandeweygaet_mst, kyrzewena_mst, itoh_mst, coles_1998_mst, adami_mst_1999, colberg_mst_2007, alpaslan_mst_2014, beuret_mst_2017, libeskind_web, Bonnaire_2020, Bonnaire_2022}. More recently, \cite{cosmo_from_graphsNaidoo_2022, naidoo_mst_2020} use the minimum spanning tree (MST) computed from the \textit{Quijote} simulations to compute the cosmological information by binning branch and shape features of the MST computed over the simulation suite. \cite{graph_for_halos22} illustrate graph-based approaches for modelling small-scale halo clustering in cosmological simulations.

The advent of deep learning in cosmology has made massive data generation and analysis more tractable. Many studies have investigated neural techniques for point estimate cosmological parameter extraction from cosmological fields via regression networks trained on simulation-parameter pairs \citep{pan2020cosmological, ravanbakhsh2017estimating, Kwon_2020, prelogovic2021machine, Fluri_2019lensing, Fluri_2018, Matilla_2020, ribli2018improved, Gillet_2019}, field reconstruction \citep{seljak_trenf, shirleyneural_emu}, foreground removal emulation  \citep{makinen2020deep21, 2022MNRAS.510L...1J}, or cosmological parameters from graphs \citep{paco_graphs} with squared loss.  As reviewed in \cite{villaescusanavarro2020neural}, these techniques can estimate the posterior mean of parameters (see also \cite{jeffrey2020solving}). This implies they require simulations drawn from a prior, specified at the time of training, not just near the parameters favored by the data. This adds to the variability that needs to be fit by the network.

We take a different approach: we consider halo catalogue graphs as our
dataset and use Information Maximising Neural Networks (IMNNs) to measure
the Fisher information contained in these graphs. IMNNs are neural networks
that compress data to informative nonlinear summaries, trained on
simulations {around a fiducial model} to maximise the Fisher information \citep{Charnock_IMNN, makinen2021}, where, for the purposes of compression and forecasting only, the summary statistics are assumed to have a Gaussian sampling distribution. Neural networks can
make use of all available data simultaneously, even saturating known
field-level likelihoods \citep{makinen2021}. This approach enables us to use asymptotically optimal nonlinear statistics \citep{alsing2018_general, Charnock_IMNN} to {then} compute summaries and estimate maximum likelihood parameters and perform efficient implicit likelihood inference over a prior. 

We combine this framework with a graph neural network (GNN) architecture. GNNs are well-suited to discrete and variable-length problems such as molecular classification, weather forecasting, and even physics (re)-discovery with symbolic regression \citep{lemos2022rediscovering, cranmer2020discovering}, (see \cite{battaglia2018relational} for a complete review).

Recent studies have made use of IMNNs for cosmology \citep{makinen2021, fluri2021, fluri_kids}, and highly non-Gaussian problems, such as galaxy type identification from multiband images \citep{livet2021catalogfree}. However, previous implementations relied on computing Fisher statistics for data with a \textit{fixed input size}. Here, using GNNs, we extend the framework to a much more general class of problems. We will refer to  graph IMNNs as gIMNNs.

We show how gIMNN summaries from catalogue graphs compares to
traditional cosmological techniques \emph{with respect to
information extraction} using the \textit{Quijote} halo
catalogues \citep{quijote2020}. We illustrate that by encoding
physical symmetries and more descriptive graph attributes in
the IMNN framework, we can extract more information from
limited catalogues than traditional 2-point statistics.

The study is organised as follows: We present a graph description of large-scale structure in Section \ref{sec:lss-graphs}, followed by a review of the IMNN framework in the context of graph data in Section \ref{sec:imnn}. In Section \ref{sec:cosmo} we present our halo catalogue graph and GNN architectures and our main findings: We first investigate
information as a function of increasing GNN depth and graph
connectivity on both invariant and non-invariant graphs, and
show that gIMNNs consistently extract more information than the
2-pt function. We next show that decorating graph nodes with
mass further increases information extraction. Third, we
explore the information stored in graph cardinality (the number
of nodes or objects and edges connecting them) in the context
of the halo mass function. Next, we proceed to a more realistic
case in which catalogue construction is subject to various
levels of uncertainty in the halo mass determination. In Section \ref{sec:sbi} we
conclude by showing how trained gIMNN summaries can be used as
optimal compressors in simulation-based inference density
estimation. We include supplementary descriptions of graph
assembly and network generalization in {Appendix
\ref{app:graph-assembly}}.

\section{Large-Scale Structure as a Graph}\label{sec:lss-graphs}
\begin{figure}[htp!]
    \centering
    \includegraphics[width=\columnwidth]{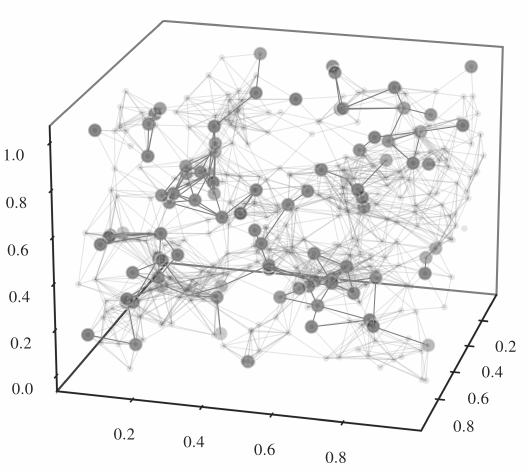}
    \caption{Dark matter halo graph representation of large-scale structure, constructed from a single \textit{Quijote} simulation. The largest halos (grey) with a mass $M_i > 1.5\times10^{15}\ M_\odot$ trace the largest physical scales, here shown coloured by the log of the halo's mass, and connected to all neighbours within a radius of $r_{\rm connect}=200\ \rm Mpc$. Smaller halo masses $M_i > 1.1\times10^{15}\ M_\odot$ (light grey) trace smaller scale clustering. The box encases a cosmological comoving volume of {$(1\textrm{Gpc})^3$}.}
    \label{fig:ghostgraph}
\end{figure}

Graphs provide a natural language with which to describe the cosmic web.
Dark matter halos are attributed to nodes (vertices), while filaments are traced by smaller halos and edges, illustrated in Figure \ref{fig:ghostgraph}. In this representation, clustering under gravitational interactions can be translated into higher edge cardinality (number of edges). Higher order $n$-point functions can be computed efficiently for clusters, while avoiding the cost of computing extraneous connections across voids. Void catalogues (where edges would correspond to the walls separating the voids) can likewise be assembled into the dual of a halo graph. Graph construction also allows arbitrary extra information, such as the halo masses or peculiar velocities, for which the underlying sampling distribution is not known, to be combined in a nonlinear fashion in the form of node or edge labels, unlike (marked) correlation functions.

\subsection{Graph Notation}\label{sec:graphs}
We define a graph explicitly as a tuple $G = (\myglobal, V, E)$, following the notation in
\cite{battaglia2018relational}. The $\myglobal$ is a
global attribute of the graph, i.e. a label or global parameter value. $V = \{ \nodei \}_{i=1:N^v}$ is the set
of graph nodes, with cardinality $N^v$. The edge set $E = \{(\edge, r_k, s_k) \}_{r_k = i, k=1:N^e}$, indexed by
$k=1:N^e$, is comprised of vectors  $\edge$ of cardinality $N^e$, which may be directed, connected via
receiving and sending {indices} between nodes, $r_k$ and $s_k$. Senders and receivers can be equivalently parameterized by an adjacency matrix $A_{ij}$ in which $i$ and $j$ index sender and receiver nodes, respectively. Each node, indexed by $i=1:N^v$, has a set of
edges, $E_i' = \{(\edge', r_k, s_k) \}_{r_k = i, k=1:N^e}$, connected to it via a subset of senders and
receivers. The full set of nodes is defined as $V= \{\nodei \}_{i=1:N^v}$, where each node $\nodei$ is a vector of features. In a physical system of particles, one
might represent $V$ as a set of individual particles' attributes, like mass, position, and velocity, with edges
expressing interactions, such as forces, between particles. A global
attribute of a graph might be a classification label, such
as in molecule or cluster classification
\citep{argmax_flow, kipf_gcn}. Careful data representation
on graphs can vastly simplify physical problems via
inductive biases and symmetry capture \citep[see e.g.][]{lemos2022rediscovering, cranmer2020discovering, battaglia2018relational}. 

\subsection{Halo Graphs}\label{sec:halographs}

We define a dark matter halo graph $G=(\myglobal, V^{\rm halo},E^{\rm halo})$, constructed from a catalogue for a
single realisation of the universe. We can equivalently define its dual, $H=( \myglobal, V^{\rm void}, E^{\rm void})$, from a void catalogue. Note that if we assign
global cosmological parameters to $\myglobal$, $G$ and $H$ share this property. Hereafter we will focus on graphs
from halo catalogues.

The graph framework allows the cardinality of a
cosmological graph's nodes and edges to vary as a function
of cosmological or survey parameters, reflecting the
often strong dependence of the abundance of clusters on cosmological parameters. When assembling a graph from a
halo catalogue, we choose to vary two physical parameters:
a mass cut, $M_{\rm cut}$, and a linking radius, $r_{\rm connect}$. A halo $i$ with a mass above $M_{\rm cut}$ is
connected to a halo $j$ if the absolute distance between halos $i$ and $j$ is less than $r_\textrm{connect}$, i.e. $|\textbf{d}_{ij}|<r_{\rm connect}$. We display the same catalogue at two mass cuts in Figure \ref{fig:ghostgraph}. A conservative $M_{\rm cut}=1.5\times 10^{15} M_\odot$ (dark points) contains the heaviest halos and traces the largest scales, while smaller masses ($M_{\rm cut}=1.1\times 10^{15} M_\odot$, light points) trace smaller scales. Each graph is connected by $r_\textrm{connect}=200\rm \ Mpc$.

Graphs can be assembled from halo catalogues in one of two ways: as non-invariant or as invariant graphs. Non-invariant
graphs have positions, $\boldsymbol{p}$, as node labels, setting $\nodei = \boldsymbol{p}_i$, with edges labelled as the relative
distances between halos, $\textbf{d}_{ij}$. This graph is \textit{not} invariant under translations and rotations, as the node
values are pinned to the underlying simulation grid. 
\textit{Invariant} graphs have only \textit{relative}
positional information, all of which is stored in the edges. The
cosmological models that we wish to constrain are invariant to rigid Euclidean group rotations and translations of the large-scale
structure. In this work we include both representations for completeness.

\subsubsection{Node features}
In the invariant representation, graph nodes are `decorated' with either an indicator $\nodei = v_i = {1}$ in the undecorated case or the halo's scalar mass, $\nodei = v_i = M_i$. In the non-invariant case, nodes are also decorated with position $\nodei = (M_i, \textbf{p}_i)$. We describe graph construction and padding details in Appendix \ref{app:graph-assembly}.

\subsubsection{Edge features}\label{sec:edgeconstruct}
To construct invariant graphs, we impose translational symmetry by
attributing functions of relative positions between
halos on the edges. We compute the vector separations 
$\textbf{d}_{ij} = \boldsymbol{p}_i - \boldsymbol{p}_j$ between all halos and do not link halos directly if $|\textbf{d}_{ij}|> r_{\rm connect}$. For rotational invariance, we adopt \cite{paco_graphs}'s notation and first compute the unit vectors $\boldsymbol{s}_{ij} = \textbf{d}_{ij} / |\textbf{d}_{ij}|$ and $\textbf{n}_i = (\boldsymbol{p}_i - \bar{\boldsymbol{p}})/ |\boldsymbol{p}_i - \bar{\boldsymbol{p}}|$ where $\bar{\boldsymbol{p}}$ is the centroid (or reference halo position). We then compute the direction cosines $a_{ij} = \textbf{n}_i \cdot \textbf{n}_j$ and $b_{ij} = \textbf{n}_i \cdot \textbf{s}_{ij}$. The normalized edge features for \textit{invariant} halo graphs are then
\begin{equation}\label{eq:edges}
    \textbf{e}_{ij} = \left[|\textbf{d}_{ij}| / r_{\rm connect}, a_{ij}, b_{ij}\right],
\end{equation}
whilst for {non-invariant} graphs, $\textbf{e}_{ij} = |\textbf{d}_{ij}| / r_{\rm connect}$.

\subsubsection{Global features}\label{sec:globalconstruct}
A halo graph's global features can be any quantity that describes the global properties of the system, in this case configuration or cosmology parameters. In a regression case, one might wish to label each halo graph simulation with a set of cosmological or hydrodynamical parameters, as done in \cite{paco_graphs}, and fit a neural network to minimise some distance measure between the network output and these parameters. Here, global properties will be \textit{arbitrary nonlinear summaries} of cosmology, learned in an unsupervised manner as a function of the graph's attributes using information maximising neural networks. 

\section{Information Maximising Neural Networks}\label{sec:imnn}

\begin{figure*}[ht!]
    \centering
    \includegraphics[width=\textwidth]{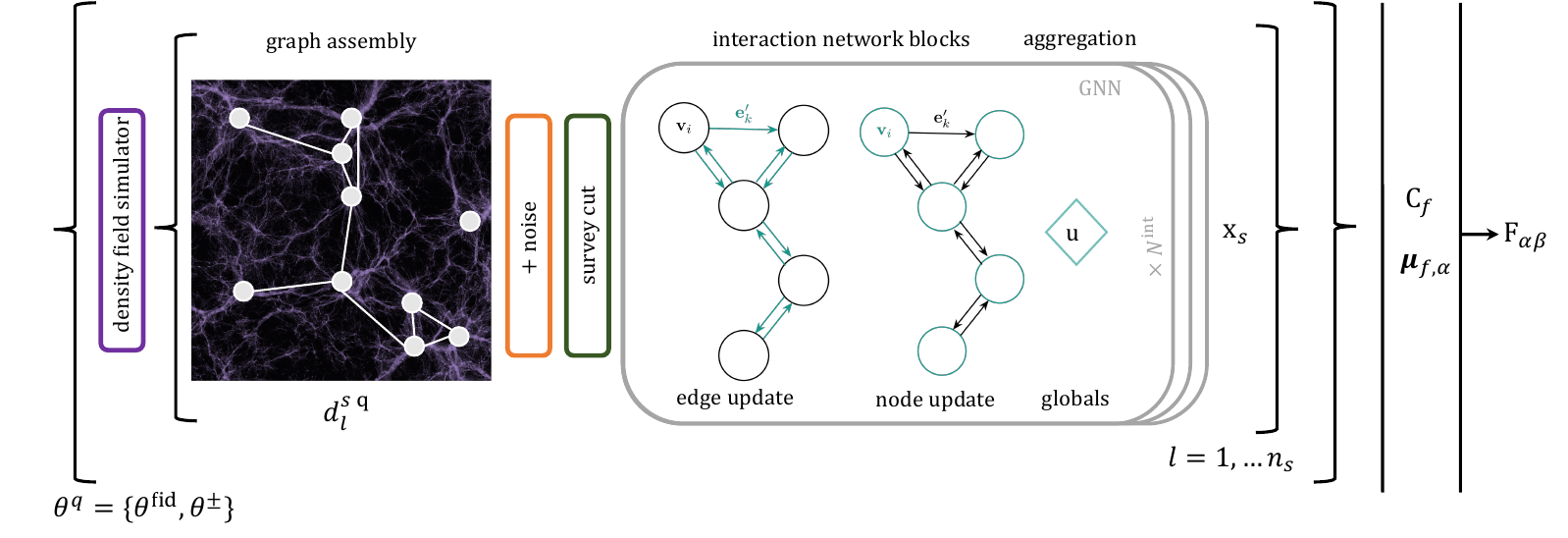}
    \caption{Cartoon of the graph-based information maximising neural network scheme. For each $i=1,\dots n_s$ dark matter field realization, halo catalogues are computed from a density field and assembled into a connected graph. A GNN block then computes edge and node updates (green) outlined in Section \ref{sec:gnns}, pooling graph attributes to compute global summaries, $\xvec = \myglobal $.  This process is repeated for simulations at the fiducial parameter values, as well as at $\parvec^\pm$ for numerical derivative calculation via equation \eqref{eq:discrete-deriv}.  The output of the IMNN is the Fisher information matrix, computed via equation \eqref{eq:fisher-compressed}. The network is trained via gradient back-propagation, with the $\det \textbf{F}$ and $\textbf{C}_f$ contributing to the scalar loss function.}
    \label{fig:graph-scheme}
\end{figure*}

The graph framework allows for a modular study of the
cosmological information embedded in large-scale structure. We
next review IMNNs as a tool for information extraction, as well
as optimal compression for graphs assembled from cosmological surveys. 
The IMNN framework is presented in full in \cite{Charnock_IMNN}{ with developmental updates discussed in }\cite{makinen2021},
but we review the formalism here for completeness and
introduce new aspects to the technique. The sharper the peak of
an informative likelihood function $\mathcal{L}(\textbf{d} | 
\parvec )$ for some fixed data $\textbf{d}$ with $n_{\textbf{d}}$ data
points and $n_{\parvec}$ parameters at a given value of $\parvec$,
the {more informative $\parvec$ is about the data}. The Fisher information matrix
describes how much information $\dat$ contains about the
parameters, and is given as the second moment of the score of the
likelihood
\begin{equation}\label{eq:fisher-cov}
    \textbf{F}_{\alpha \beta} = \int d \dat\ \mathcal{L}(\dat | \parvec) \frac{\partial \ln \mathcal{L}(\dat | \parvec)}{\partial \theta_\alpha} \frac{\partial \ln \mathcal{L}(\dat | \parvec)}{\partial \theta_\beta},
\end{equation}
and can be written as
\begin{equation}\label{eq:fisher-def}
    \textbf{F}_{\alpha \beta} = -\left\langle \frac{\partial^2 \ln \mathcal{L}}{\partial \theta_\alpha \partial
\theta_\beta} \right\rangle \Big|_{\parvec = \parvec_{\rm fid}},
\end{equation}
evaluated at some fixed fiducial parameters.
A large Fisher information for a set of data indicates that the data is very informative about the model
parameters attributed to it. Fisher forecasting for a given model is made possible by the information inequality and the Cram\'er-Rao bound \citep{cramerharald_1946, rao_1945}, which states that the minimum variance of the value of an estimator $\parvec$ is given by 
\begin{equation}\label{eq:cramer-rao}
    \langle (\theta_\alpha - \langle \theta_\alpha \rangle ) (\theta_\beta - \langle \theta_\beta
    \rangle) \rangle \geq \textbf{F}^{-1}_{\alpha \beta}.
\end{equation}

We will write the compression as a function $f:\dat \rightarrow \xvec$. For large datasets, data compression is essential for inference to avoid the curse of dimensionality. The MOPED formalism \citep{Heavens_2000} gives optimal score compression for cases where the likelihood and sampling distributions are exactly Gaussian.

IMNNs are neural networks that perform data compression and compute the Fisher information of a data set. Such compression is possible even if the data likelihood is unknown or intractable, simply based on having simulations of the data at a given fiducial parameter point and local information about how the parameters change the data distribution. It can be shown (Wandelt, 2022, in preparation) that the optimality of the IMNN summaries holds for any unknown or intractable data likelihood even though the IMNN maximizes Fisher information assuming the {parameter-independent covariance form of the }Gaussian likelihood for the IMNN summaries
\begin{equation}\label{eq:og-moped-like}
    -2 \ln \mathcal{L}(\xvec | \dat) = (\xvec - \muvec_f(\parvec))^T \textbf{C}_f^{-1}(\xvec - \muvec_f(\parvec))
\end{equation}
where 
\begin{equation}
    \muvec_f(\parvec) = \frac{1}{n_s} \sum_{i=1}^{n_s} \xvec_i^s
\end{equation}
is the mean of the compressed summaries $\textbf{x}_i^s$, with $\{\textbf{x}^s_i | i \in [1, n_s] \}$' and we assume a parameter-independent covariance matrix. Here $i$
indexes the random initialisation of $n_s$ simulations,
and the superscript $s$ denotes {quantities derived from simulations, unlike quantities without the superscript $s$ which are derived from actual observations}.
The {summaries} are obtained via simulation {of data} $\dat^s_i = \dat_i^s(\parvec, i)$ via the compression scheme
$f:\dat_i^s \rightarrow \textbf{x}_i^s$. The covariance of
the summaries is computed from the data as well:
\begin{equation}\label{eq:imnn-cov}
    (\textbf{C}_f)_{\alpha \beta} = \frac{1}{n_s - 1}\sum_{i=1}^{n_s} (\xvec_i^s - \muvec_f)_\alpha (\xvec_i^s - \muvec_f)_\beta.
\end{equation}
{Note that this covariance is assumed to be independent of
the parameters, which, whilst not strictly true, is
enforced by regularisation during the fitting of the IMNN.}
A Fisher matrix can then be computed from the likelihood in equation \eqref{eq:og-moped-like}:
\begin{equation}\label{eq:fisher-compressed}
    \textbf{F}_{\alpha \beta} =  \tr [\muvec_{f,\alpha}^T \textbf{C}^{-1}_f \muvec_{f, \beta}],
\end{equation}
where we introduce the notation {$\bm{y}_{,\alpha}\equiv{\partial \bm{y}}/{\partial\theta_\alpha}$} for partial derivatives
with respect to parameters. If the compression function
$f$ is a neural network parameterized by layer weights $\textbf{w}^\ell$ and biases
$\textbf{b}^\ell$ (with $\ell$ the layer index), the summaries (and respective mean and covariance) then become
functions of these new parameters $\xvec(\parvec) \rightarrow \xvec(\parvec, \textbf{w}^\ell, \textbf{b}^\ell)$. To evaluate equation \eqref{eq:fisher-compressed} for a neural compression, we must compute
\begin{equation}
    \muvec_{f,\alpha} = \frac{\partial}{\partial \theta_\alpha} \frac{1}{n_s}\sum^{n_s}_{i=1} \xvec^{s\ \rm fid}_i.
\end{equation}
One way of computing the derivatives of the summary means with respect to the parameters is to define a finite difference gradient dataset by altering simulation fiducial values by a small amount, yielding
\begin{equation}\label{eq:discrete-deriv}
    \left( \frac{\partial \hat\mu_i}{\partial \theta_\alpha} \right)^{s\ \rm fid} \approx \frac{1}{n_s}\sum^{n_s}_{i=1} \frac{\xvec^{s\ \rm fid +}_i - \xvec^{s\ \rm fid -}_i}{\Delta \theta^+_\alpha - \Delta \theta^-_\alpha}.
\end{equation}
{To prevent extra information being extracted from accidental correlation in limited sized data sets,} reported statistics need to be computed on a validation set of simulations, which is unlikely to share the same accidental correlations as the fixed training set.
An alternative explored in \cite{makinen2021} is to calculate the adjoint gradient of the simulations as well as the derivatives of the network parameters with respect to the simulations:
\begin{equation}\label{eq:autograd-enabled}
    \muvec_{f,\alpha} =  \frac{1}{n_s}\sum^{n_s}_{i=1} \left( \frac{\partial \xvec}{\partial \theta_\alpha} \right)_i^{s\ \rm fid} = \frac{1}{n_s} \sum^{n_s}_{i=1} \sum^{n_d}_{k=1} \frac{\partial \xvec^{s\ \rm fid}_i}{\partial d_k}\frac{\partial \dat^{s\ \rm fid}_i}{\partial \theta_\alpha} .
\end{equation}
If the gradient of the simulations can be computed efficiently, this technique for computing the compression Fisher information eliminates the need for hyperparameter tuning of the finite difference derivative size, $\Delta \theta_\alpha$.

The network is trained to maximise the logarithm of the determinant of the Fisher information, computed via equation \eqref{eq:fisher-compressed}. 
As described in \citeauthor{Charnock_IMNN} and \citeauthor{livet2021catalogfree}, the Fisher information is invariant to nonsingular linear transformations of the summaries. To remove this ambiguity, a term driving covariance to the identity matrix is added
\begin{equation}
    \Lambda_C = \frac{1}{2} \left( \left|\left|(\textbf{C}_f-\mathbf{1})\right|\right|^2_\mathcal{F} + \left|\left|(\textbf{C}^{-1}_f-\mathbf{1})\right|\right|^2_\mathcal{F} \right),
\end{equation} 
where  $||\bm{A}||_\mathcal{F}\equiv \sqrt{\tr{\bm{AA}^T}}$ denotes the Frobenius norm. This yields the loss function
\begin{equation}\label{eq:imnn-loss}
\Lambda = -\ln \det \textbf{F}  + r_{\Lambda_C} \Lambda_C,
\end{equation}
with regularization parameter
\begin{equation}
    r_{\Lambda_C} = \frac{\lambda \Lambda_C}{\Lambda_C + \exp(-\alpha \Lambda_C)},
\end{equation}
where $\lambda$ and $\alpha$ are user-defined parameters. When the covariance is far from identity, the $r_{\Lambda_C}$ function is large and the optimization focuses on bringing the
covariance and its inverse back to identity. The network is trained until the Fisher information stops increasing for a pre-determined number of iterations.
We stress that the value of $\textbf{F}$ reported as an information metric, however, is the one computed via Eq. \ref{eq:fisher-compressed}, computed over a validation set of simulations in the case of a finite set of data. 

To summarise the IMNN algorithm we take the following steps \textit{every training epoch} to optimise the Fisher information:
\begin{enumerate}
    \item[i)] compress simulations at the fiducial model with different random seeds to the network. Calculate the covariance of these summaries using equation \eqref{eq:imnn-cov}.
    \item[ii)] compress simulations generated at perturbed fiducial parameter values, $\boldsymbol{\theta}^\pm$ to produce $\textbf{x}^\pm$. Calculate the derivatives $\textbf{x}_{f,\alpha}$ with equation \eqref{eq:discrete-deriv}. 
    \item[iii)] Calculate the Fisher matrix (Eq. \eqref{eq:fisher-compressed}). Pass the Fisher and covariance matrices to the loss function. Update neural network weights using gradient descent such that $ \det \textbf{F}$ increases.
\end{enumerate}

\subsection{Graph Neural Networks}\label{sec:gnns}
A graph neural network (GNN) block typically consists of three update functions, {$\bm{\phi}=(\phi_u, \phi_v, \phi_e)$}, and three aggregation functions, {$\bm{\rho} =(\rho^u, \rho^v, \rho^e)$}, applied sequentially to a graph tuple  $G = (\myglobal, V, E)$. 
A single graph block $\ell$ is comprised of several update steps to its elements:
\begin{enumerate}
    \item \textit{Edge update}: Each edge is parameterized by a function $\phi^{\ell+1}_e$ which takes as inputs its connected nodes, previous value, and graph global properties and yields another edge:
    \begin{equation}\label{eq:edgeupdate}
        \textbf{e}_{ij}^{\ell + 1} = \phi^{\ell + 1}_e(\node_i^\ell, \node_{j}^\ell, \textbf{e}_{ij}^\ell \myglobal^\ell),
    \end{equation}
    where $\node_{i}^\ell$ and $\node_{j}^\ell$ are sender and receiver nodes indexed by $(s_k, r_k)$.
    \item \textit{Node update}: Each node is then parameterized by a function $\phi^{\ell + 1}_v$ and outputs a new node:
    \begin{equation}\label{eq:nodeupdatealt}
        \nodei^{\ell + 1} = \phi^{\ell+1}_v \left(\rho^{e \rightarrow v}(E_i^{\ell + 1}), \nodei^\ell, \myglobal^\ell \right),
    \end{equation}
     Here a permutation-invariant aggregation operation $\rho^{e \rightarrow v}(E_i^{\ell + 1})$ pools the neighbourhood of edges $E^{\ell + 1}_i$ connected to node $i$ into a fixed-sized vector to feed into the update function.
    \item \textit{Global update}: The global features of the graph are then updated with a function $\phi^{\ell + 1}_u$:
    \begin{equation}\label{eq:globalupdatealt}
         \myglobal^{\ell + 1} = \phi^u\left(\rho^{e\rightarrow v}(E^{\ell + 1}), \rho^{v \rightarrow u}(V^{\ell + 1}), \myglobal^\ell \right),
    \end{equation}
    where the graph's edge ($E^{\ell+1}$) and node ($V^{\ell+1}$) sets are pooled into fixed-sized vectors for the global update. 
\end{enumerate}
The order of operations of these updates is flexible, but usually applied in the order displayed above, and in the GNN block in Fig. \ref{fig:graph-scheme}. This framework allows {$\bm{\phi}$} functions to be arbitrarily parameterized as neural networks with nonlinear activation functions. Aggregation functions {$\bm{\rho}$} must be allowed to take a variable number of arguments, so are usually chosen to be permutation-invariant operators such as the mean, summation, or maximum \citep{bronstein2021_permeq}. Stacking $\ell = 1:N^{\rm int}$ GNN blocks allows node information to be propagated to and from neighbours $N^{\rm int}$ degrees away, where $\textrm{int}$ refers to \emph{interactions}. In this work all GNN blocks operate over the entire graph. However, one could also devise surrogate GNN blocks that operate on small scales and then pass information up to larger scales via an aggregation function $\rho^{\rm small \rightarrow large}$, such that one GNN network is not responsible for operating on nodes of all scales in a densely-populated graph. We detail our specific implementation and architecture in Section \ref{sec:architecture}.

The GNN framework is readily incorporated into the IMNN formalism, since the details of the neural network architecture only serve to better capture how the data changes with the parameters. Instead of predicting an output graph or class label, as in \cite{battaglia2018relational}, our final global update $\phi^{u}$ outputs IMNN summaries, $\xvec = \myglobal^{\ell = N^{\rm int}}$. This new aspect to the IMNN formalism is the ability to operate over variable-length data inputs, rendering the cardinality of input graphs, $n_{\rm data} = N^v$ and $N^e$ informative features of the data. A stochastic system might yield a different number of discrete particles for different parameters, meaning the \textit{number} of data becomes a descriptor of the statistical model. This allows for a study of information $\mathcal{I}= \frac{1}{2}\ln \det F$ as a function of $N^v$ and enables much more flexible data modelling.

\section{Cosmological Parameter Inference with Halo catalogues}\label{sec:cosmo}
\begin{figure}[htp!]
    \centering
    \includegraphics[width=\columnwidth]{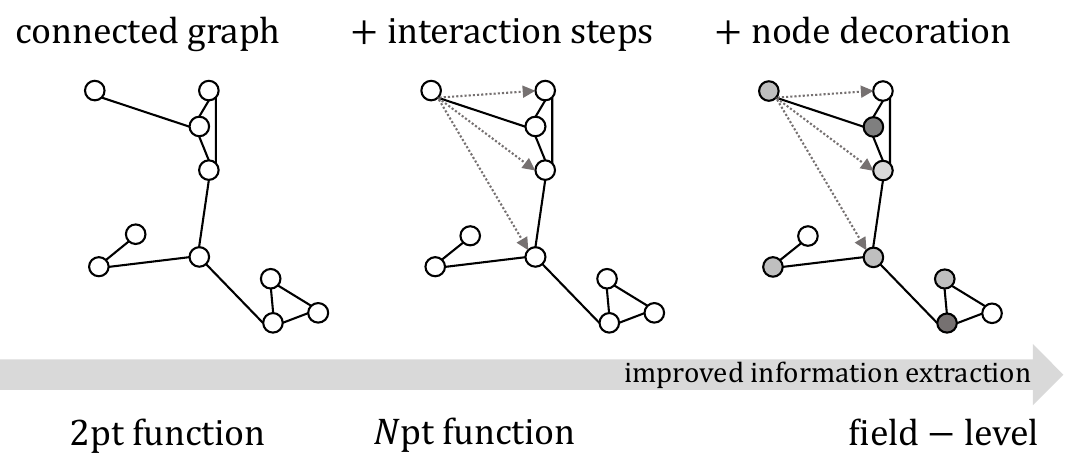}
    \caption{Information as a function of large-scale structure data representation. Cosmological parameter information extraction efficiency increases as information is added and propagated throughout the halo graph with increased connections and message-passing steps. The second and third graph show message propagation from the top left node after $N^{\rm int}=2$ update steps. }
    \label{fig:info_diagram}
\end{figure}
Here we consider applications of our graph IMNNs on realistic
cosmological problems. Even future astronomical studies will not
be able to image complete dark matter overdensity fields of
large-scale structure. However, discrete galaxy and void catalogues
can be assembled as tracers of structure. Different LSS realizations from stochastic initial conditions will have different numbers of halos
and voids, posing a problem for usual fixed-size neural
networks, {(which are themselves problematic for inference unless treated correctly)}.
We explore information extraction as a function of graph
connectivity in the context of two-parameter inference for
the matter density parameter, $\Omega_m$, as well as
$\sigma_8$, the r.m.s. fluctuation of density perturbations at the
$8\ \rm h^{-1}\ Mpc$ scale. Both {$\Omega_m$ and $\sigma_8$} parameterize the distribution of
matter in cosmological simulations, so the graph topology should be sensitive to
changes in parameters.

The more descriptive a graph is, the more information one intuitively expects to extract. We demonstrate this trend by first considering \textit{undecorated} graphs, annotated with just positions of and relative distances between halos. We show that information extraction efficiency increases as graph connectivity increases. We then show that information increases further when halo masses are included as node features.

The closest existing statistics to this representation are $n$-point correlation and mass functions, and we show how information increases beyond the 2-point correlation function with the same catalogue as graph connectivity is increased, as illustrated in the cartoon in Figure \ref{fig:info_diagram}.

\subsection{Halo Catalogues}

Here we describe the simulated catalogues that are used
for training and validation.The Quijote Halo catalogues
are assembled from 3D overdensity fields at the present
day ($z=0$) using the Friends of Friends (FoF) algorithm
\citep{davis_fof}. Attributes computed by the finder are
halo masses $M_i$, positions $\boldsymbol{p}_i$, and
velocities $\boldsymbol{v}_i$. Each full simulation yields
a catalogue of $\sim$400,000 halos on average. Here we restrict our analysis to mass and clustering information in an effort to compare our method to known statistics.

\subsubsection{Graph inputs assembly}
We initially connect graphs of a manageable size by varying two hyperparameters. We first make a minimum mass cut $M_{\rm cut}$ to be considered in the catalogue. Nodes are then connected to one another within a Euclidean distance $r_{\rm connect}$. We initially explore the noise-free limit with known masses and fix $M_{\rm cut}=1.5\times10^{15}M_\odot$ to assess the pure information limit of the catalogues. We add noise to the analysis in Section \ref{sec:withnoise}. This cut yields $N^v \in [70, 140]$ halos per catalogue. We visualize two graphs in Appendix \ref{app:graph-assembly}, figure \ref{fig:graph-scheme}. 

We then assemble the truncated catalogues into non-invariant and invariant graphs, as outlined in Section \ref{sec:halographs}. We initialize each graph's global property with a tuple $\myglobal^{\ell=0} =(\arcsinh{N^v}, \arcsinh{N^e})$ summarising the cardinality of the graphs. As described in \cite{battaglia2018relational, lemos2022rediscovering, paco_graphs}, imposing symmetries in data representation can improve GNN training, since the network can focus on learning
relevant correlations to the problem, as opposed to
re-learning symmetry. We test this notion in the context of information extraction. 

\subsubsection{Graph neural network architecture}\label{sec:architecture}
We choose to parameterize our GNN functions with simple fully-connected networks. Each $\bm{\phi}$ function is a dense network with two layers of 50 hidden neurons and \texttt{gelu} activations \citep{gelu_ref}. We built a custom aggregation function akin to that found in \cite{paco_graphs}, in which mean, max, sum, and variance are computed over node and edge attributes and then concatenated, since it is not known a priori which function is most useful for information extraction. To aggregate e.g. the set of edges $E_i$ in a neighbourhood around node $i$ we compute:
\begin{equation}\label{eq:customagg}
     \bigoplus_{j \in E_i} \textbf{e}_{ij} = \left[\max_{j \in E_i} \textbf{e}_{ij}, \sum_{j \in E_i} \textbf{e}_{ij}, \frac{\sum_{j \in E_i} \textbf{e}_{ij}}{\sum_{j \in E_i} } \right]
\end{equation}
We additionally modify these operators with a trainable $\arcsinh$ layer e.g. for edge-to-node aggregation: 
\begin{equation}
    \rho^{\ell+1}_{e\rightarrow v}(E^\ell_i) = a\arcsinh\left(b\bigoplus_{j \in E^\ell_i} \textbf{e}^\ell_{ij} + c\right) + d,
\end{equation}
where $(a,b,c,d)$ are scalar learnable parameters initialized as $(1,1,0,0)$ to ensure numerical stability for gradient calculation. All networks are trained with an Adam optimizer with a learning rate set to $0.0001$ and coupling parameters $\lambda=10$ and $\alpha=0.95$. We construct our graphs and GNNs using the \texttt{jraph} \citep{jraph2020github} and \texttt{Flax} \citep{flax2020github} libraries, which are both $\texttt{Jax}$-compatible.

We train our gIMNNs by splitting the \textit{Quijote} simulations into equally-sized training and validation sets. Gradient descent is performed on training data, while reported compression statistics (Fisher information) are computed for the \textit{validation} set using equation~\eqref{eq:fisher-compressed}. Both training and validation sets comprise of $n_s=500$ fiducial simulations at $\parvec_{\rm fid} = (\Omega_m, \sigma_8) = (0.3175, 0.834)$ and $n_d=250$ seed-matched derivative simulations perturbed by $\delta \parvec = (0.01, 0.015)$, yielding $n_d \times 2 \times 2=1000$ simulations (see \cite{quijote2020} for details). Training on the loss defined in Eq \ref{eq:imnn-loss} is performed until a patience criterion is met, in this case, when the training Fisher information stops increasing significantly for 1000 epochs.

\subsection{Undecorated Graphs vs. $n$-point Statistics}
We first consider an undecorated graph representation of halo catalogues \textit{without} descriptive node features. Drawing more edges between nodes increases the connectivity of the graph, allowing information from a single node to reach more distant neighbours. Undecorated graphs of increasing connectivity are analogous to traditional $n$-point statistics computed for galaxy catalogues. 3-point statistics for example consider triangular groupings of galaxies, and generally offer tighter constraints from large-scale structure data than the 2PCF, as shown in \citep{bispecHahn_2020}. We additionally explore invariant and non-invariant graph structures, outlined in Section \ref{sec:halographs}.
\begin{figure*}[ht!]
    \centering
    \includegraphics[width=\textwidth]{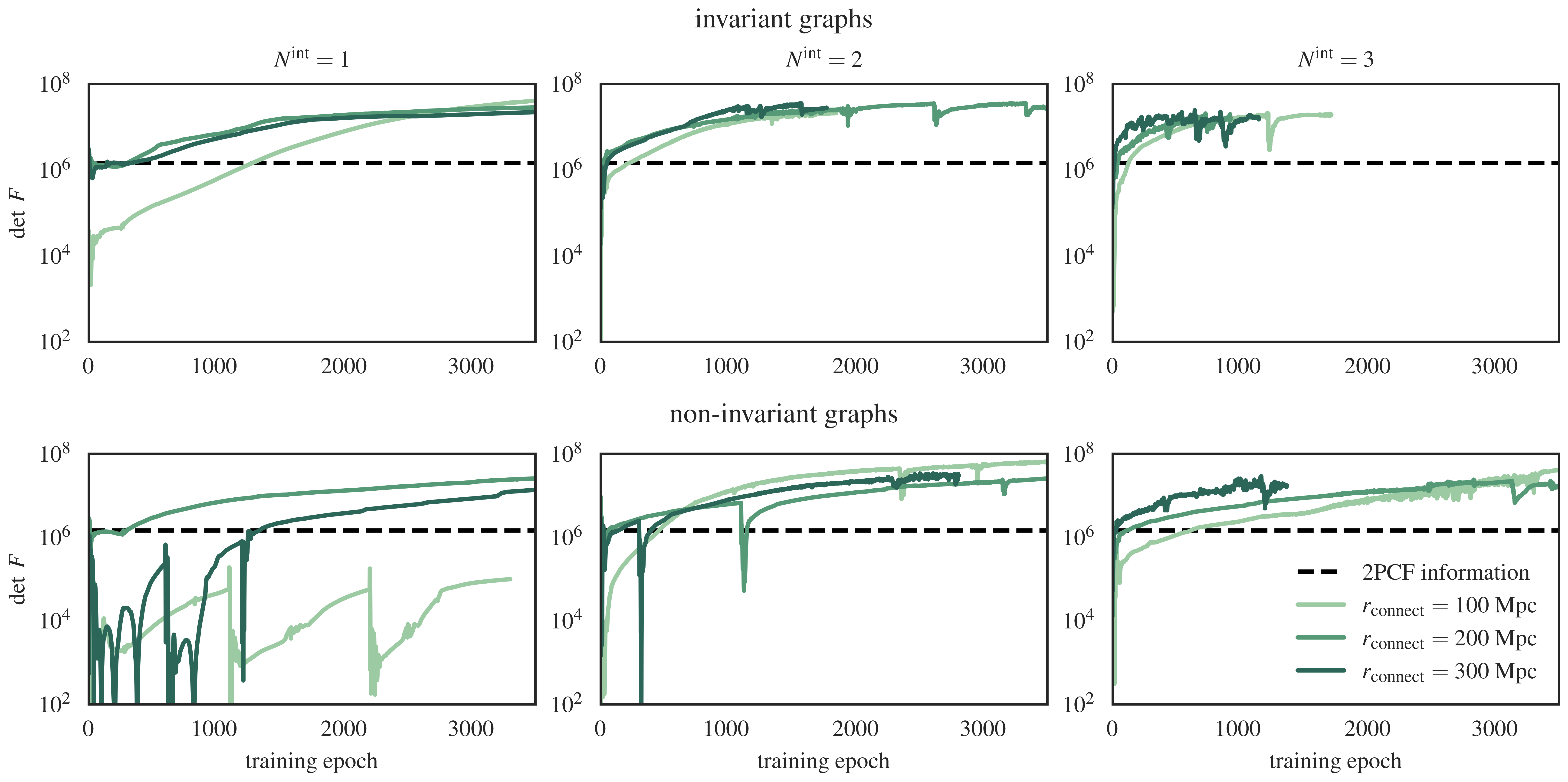}
    \caption{Information extraction comparison for undecorated invariant (\textit{top row}) and non-invariant (\textit{bottom row}) halo graphs over validation simulations. Columns indicate how many graph network update steps are performed, and colours indicate graphs assembled with different physical connection radii. The information obtained from the 2PCF computed from the same catalogues with the same mass cut is shown as a dashed line. For sufficiently descriptive networks ($N^{\rm int}>1$), invariantly-assembled graphs train faster, but all graphs connected for $r>100\ \rm Mpc$ saturate at $\det F \approx 2.8\times 10^7$, or $\approx 15$ times higher than the 2PCF. Jagged dips in validation curves indicate points of restarting network training after patience criteria were met.}
    \label{fig:connect-compare}
\end{figure*}

\vspace{0.5cm}
\subsubsection{Comparison to 2-point correlation information}
As a benchmark for our analysis, we also compare the information content obtained from the 2-point correlation function (2PCF), $\xi(r)$, of our small halo catalogues, the real-space equivalent to the \textit{Quijote} power spectrum computed in \cite{quijote2020}. For a statistic $Q=\xi(r)$, the Fisher information is given by \citep{Tegmark_1997}:
\begin{equation}\label{eq:corr-info}
F_{ij} = \frac{1}{2} {\rm tr} \left\{ \boldsymbol{C}^{-1}\left[ \left( \frac{\partial Q}{\partial \theta_i} \frac{\partial Q}{\partial \theta_j}^T \right) + \left( \frac{\partial Q}{\partial \theta_i}^T \frac{\partial Q}{\partial \theta_j} \right) \right] \right\},
\end{equation}
where $\boldsymbol{C}$ is estimated from simulations at the fiducial and the derivatives are approximated numerically via
\begin{equation}
    \frac{\partial Q}{\partial \theta_i} \approx \frac{Q(\theta^+_i) - Q(\theta^-_i)}{\theta^+_i - \theta^-_i} 
\end{equation}
We use the full suite of $500$ derivative and $1000$ fiducial halo catalogue simulations to compute Eq \ref{eq:corr-info}, and crucially \textit{make the same conservative mass cut}. We bin distances into 10 fixed bins between 0 and $\sqrt[3]{3}\rm \ Gpc^3$, yielding a covariance matrix of size $100^2$. For the 2PCF we obtain a Fisher information of $2.28\times 10^6$, or Shannon entropy of $7.319$ nats.

\subsubsection{Increasing graph connectivity}
Here we construct graphs of varying edge cardinality by varying a physical connection parameter, $r_{\rm connect} \in \{ 100, 200, 300 \}\ {\rm Mpc }$, yielding graphs with average edge number $\langle N^e \rangle \in \{27, 150, 437 \}$ respectively. We also compare information as a function of increasing GNN interaction blocks, $N^{\rm int}$, for all $r_{\rm connect}$, for both non-invariant and invariantly-structured halo graphs. Network architecture is identical for each $r_{\rm connect}$ value, and initialized by the same random seed.

\noindent \textbf{Results.} We display information extraction as a function of graph connectivity in Figure
\ref{fig:connect-compare}, computed for the validation
simulation set. We also display the catalogue's 2PCF
information (dashed line) as a benchmark. Invariant graphs
({top row}) train more smoothly than non-invariant graphs
({bottom row}) since the network does not have to learn
relationships from position values on the nodes. A single
GNN block struggles to extract information with $r_{\rm
connect}=100\ \rm Mpc$ in the non-invariant
representation (lower left), but plateaus at
$\approx2.8\times 10^7$ for all other configurations. This
behavior is likely because in most cases the network is both
descriptive enough and is able to capture patterns at much larger scales by attending to halos higher degrees away. The
common saturation value across multiple network and
connectivity combinations indicates that undecorated graphs
typically contain $\det F_{\rm IMNN} / \det F_{\rm 2PCF}\approx 10-15$ times more information than
the 2PCF, regardless of connectedness,
provided a descriptive enough network can extract it. The graph representation improves marginal constraints in $\Omega_m$ by a factor of $\approx 2$ and in $\sigma_8$ by a factor of $\approx 11$, displayed in Figure \ref{fig:F-withmass}.

We initially hypothesized that increasing network complexity would increase the information extraction. However, we found
that we obtain essentially equivalent information for any
combination of $r_{\rm connect}$ and $N^{\rm int}$. It is
clear that for this small number of halos the information is
easily saturated at any level, but with more halos in the
catalogue, as discussed in Section \ref{sec:discussion},
hierarchical clustering at the graph or network level might
pull out more information from e.g. smaller mass scales. This exploration is reserved for a future work.

Since all sufficiently-connected representations obtain the same
information, we proceed in our experiments with invariant
graphs connected with $r_{\rm connect}=200\ \rm Mpc$ and
$N^{\rm int}=2$, since this combination resulted in the
smoothest and fastest training (4 minutes) on a single
NVIDIA-v100 GPU.

\subsection{Decorated Graphs: Incorporating Halo Mass}\label{sec:decorated-graphs}
 We next decorate each halo node with the corresponding (noise-free) halo mass. We widen the network's hidden dimension to 64 and train both decorated and undecorated graphs with the same patience settings. Training was restarted for each three times after plateau to ensure saturation. The same network architecture is able to extract 2.3 times more information when decorated with masses, corresponding to 42 times more information than the 2PCF. We display the corresponding Fisher ellipses in Figure \ref{fig:F-withmass}, along with isomass lines of the halo mass function, described in Section \ref{sec:hmf-comp}. We also decorated graphs with peculiar velocities with slight improvement in information extraction but restricted our analysis to mass and clustering for interpretability.

\section{Mass cut information}\label{sec:masscut}
We next investigate how much information is contained in the mass cut, $M_{\rm cut}$ which determines halo number $N^v$. We compare two graph assemblies: one with a \textit{fixed} number of the most massive halos $N^v_\textrm{fixed}=105$, i.e. the average number of halos across variable sized halo catalogues with fixed mass cut $M_\textrm{cut} = 1.5\times 10^{15}M_\odot$, and another where the cardinality is allowed to vary with $M_{\rm cut}$. For each case we compare decorated and undecorated graphs, and the network (epistemic) and data sampling (aleatoric) errors associated with each representation. 
\begin{table}[ht!]
\begin{center}
\adjustbox{width=\columnwidth}{%
 \begin{tabular}{l l r r r}
 \toprule
 catalogue {$N^v$}\ \ \  & graph assembly\ \ \  &  {$\ln \det F$} & vary network & \textbf{vary data} \\
 \hline
     & without mass &  & $5.03\pm0.47$ & $\boldsymbol{5.98\pm1.06}$ \\
    fixed & with mass &  & $12.43\pm1.44$ & $\boldsymbol{12.39\pm0.22}$ \\
     & 2PCF & 9.74 & &  \\
\hline
     & without mass &  & $17.89\pm0.33$ & $\boldsymbol{17.66\pm0.27}$ \\
    variable & with mass &  & $17.40\pm0.57$ & $\boldsymbol{17.85\pm0.12}$  \\
     & 2PCF & 14.19 &  &  \\
 \toprule
\end{tabular}
}
    \caption{\begin{justify}
    {\textnormal{Comparison of extracted information from fixed- and variable-length catalogues. Information was extracted with $r_{\rm connect} = 200\ \rm Mpc$, and networks with $N^{\rm int}=2$ across 5 identical initialisations. We display each configuration's best Fisher information and the means and standard deviations over 5 network initializations (second-to-last column) and 5 different train/validation set splits (last column).}}
    \end{justify}
    }
\label{tab:masscut}
\end{center}
\end{table}
To estimate {network variability} we train five gIMNNs with $N^{\rm int}=2$ with different initialization of network parameters, whilst fixing the training and validation sets, displaying the best Fisher obtained as well as the mean and standard deviation of $\ln \det F$ over the five runs. For data sampling uncertainty we fix the gIMNN weight values on initialization and train on five different randomised train-validation equal-sized splits of the available simulations. For both data and network error cases, the same five random seeds and network architecture is used across all data configurations.

\noindent \textbf{Results}. We display results in Table
\ref{tab:masscut}. Fixing catalogue size eliminates halo number
as a useful feature to the network, evidenced by much lower
information yields. Without mass decoration there is less
information in the graph data so the data can be fit by more
possible functions by the network, so the variability of the Fisher as a function of the data sampling is increased over the decorated case. However, the network has to fit a simpler compression since
there are fewer relevant features without mass, so the variability in possible network weights decreases,
compared to the decorated case. Fixed-length
graphs do not exceed the 2PCF information until annotated with
mass information.

When catalogues are allowed to vary with a physical mass cut,
much more information can be extracted from both decorated and
undecorated graphs. Including mass information on the nodes
again increases the variability incurred across different
network initializations, but decreases the aleatoric
uncertainty since we better describe the likelihood with more
information.

\subsection{Comparison to the Halo Mass Function}\label{sec:hmf-comp} 

The results of Section \ref{sec:masscut} indicate that
catalogue information extraction is extremely sensitive to a
physical mass cut. This behaviour is akin to constraints
obtained using halo mass cumulative distribution functions
\citep{halomass_cosmo, countsincells_fisher, Artis_clusters_2021}. The halo number density function is $dn/dM$, defined as the
number of halos of mass $M$ per unit volume per unit interval in $M$,
equivalently parameterized using the smoothed r.m.s. linear
overdensity of the density field, $\sigma(M)$, via the halo mass function (HMF), $f(\sigma)$.
The fraction of mass in collapsed halos per unit interval $\ln \sigma^{-1}$ obeys
\begin{equation}
    \int_{-\infty}^\infty f(\sigma) d \ln \sigma^{-1} = 1,
\end{equation}
and is related to the halo number density function via
\begin{equation}\label{eq:hmf}
    \frac{dn}{dM} = \frac{\rho_o}{M}\frac{d\ln \sigma^{-1}}{dM} f(\sigma),
\end{equation}
where $\rho_o$ is the mean mass density of the universe. The form of $f(\sigma; \parvec)$ can be related analytically to cosmological parameters, such as in \cite{press_schechter_fn}, or approximated using simulations \citep{halomass_cosmo}.

We compare gIMNN Fisher constraints to isomass contours of the {integrated} Press-Schechter HMF,
$f_{\rm PS}(\sigma;\ \Omega_m, \sigma_8)$, integrated from a fixed
$M_{\rm cut}$ as a function of cosmological parameters in
Figure \ref{fig:F-withmass}. We use the HMF since this quantity incorporates both halo number and mass information. See Appendix \ref{app:hmf-comp} for a detailed comparison of $dn/dM$ and $f(\sigma)$ functions. We utilize \texttt{hmf calc} \citep{hmf_calc_murray, hmfcalc_soft} for the calculation. The HMF and the corresponding halo number density at fixed $M=M_{\rm cut}$ determines a

\begin{figure}[htp!]
    \centering
    \includegraphics[width=\columnwidth]{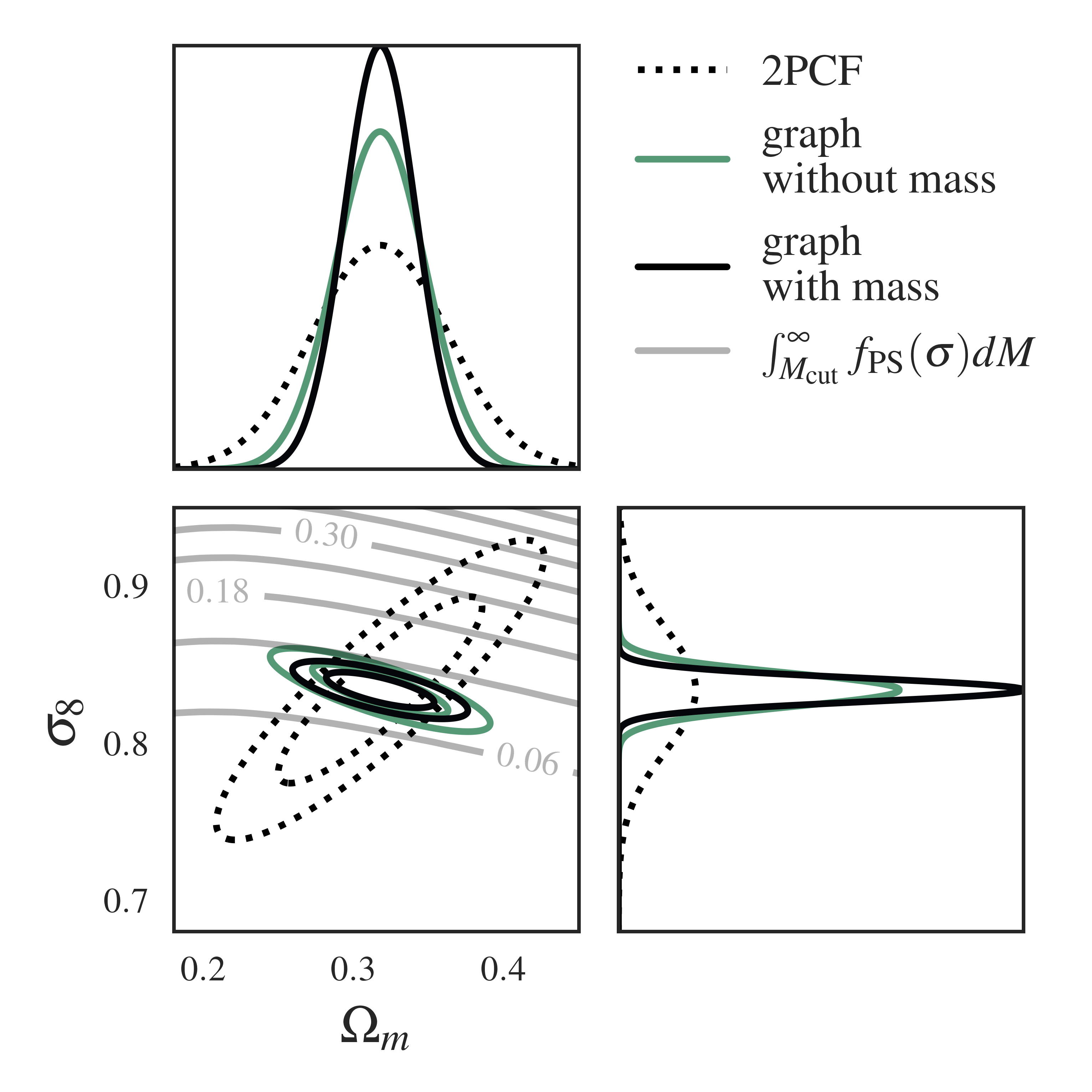}
    \caption{Fisher matrix comparison for decorated (black) and undecorated (green) graphs and 2PCF (dashed) computed for the same catalogue, plotted over contours of the Press-Schechter halo mass function (grey), where the numbers indicate the integrated mass density of halos from $M_{\rm cut}$ to infinity, as a function of $\Omega_m$ and $\sigma_8$. The HMF contours indicate the information provided by a fixed physical mass cut, orthogonal to the positional information in the 2PFC. The IMNN Fisher ellipses follow these contours, but are not degenerate in the direction of the 2PCF, indicating that the IMNN has learned to use both mass and clustering information. Graphs annotated with halo mass tighten constraints by a factor of 2.3 over undecorated graphs, and by a factor of 42 over 2PCF information.}
    \label{fig:F-withmass}
\end{figure}

relatively narrow locus in the $(\Omega_m, \sigma_8)$ plane. The clustering information, traced by the 2PCF
Fisher, (which is accessible by the network) serves to lift this degeneracy. The network Fishers are nearly
parallel to the HMF isomass contours, but are not
degenerate in the direction of the 2PCF Fisher's major
axes. Decorating nodes with masses also induces a slight
rotation towards the isomass contours, since the network
has more detailed mass information to work with. This
result indicates that \emph{the network has automatically
learned to extract information from both halo clustering and mass information}.

\section{Working with Noisy catalogues}\label{sec:withnoise}
We next add observational noise and catalogue cuts on-the-fly during gIMNN training to mimic survey assembly with imperfect observations. Before a graph is constructed from a halo catalogue and fed to the network in training, halo masses are subjected to white noise with fixed variance,
\begin{equation}
    \hat{m}_i = m_i + \mathcal{N}(0,\sigma^2_{\rm noise}),
\end{equation}
where $\sigma_{\rm noise}=A_{\rm noise}M_{\rm cut}$. Observed
halos that fall below $M_{\rm cut}$ are then trimmed from the
graph to mimic real catalogue cuts in the presence of noisy
mass estimates. This noise model reflects uncertainty in the
halo finder or galaxy catalogue builder. Smaller masses close
to $M_{\rm cut}$ are more likely to be cut due to mass
underestimation, similar to low-brightness clusters in sky
surveys. We choose $M_{\rm cut}=1.5\times10^{15}M_\odot$ and
train identical networks with different amplitudes of
on-the-fly noise; $A_{\rm noise} \in \{ 0.05, 0.1, 0.2 \}$.

\begin{figure}
    \centering
    \includegraphics[width=\columnwidth]{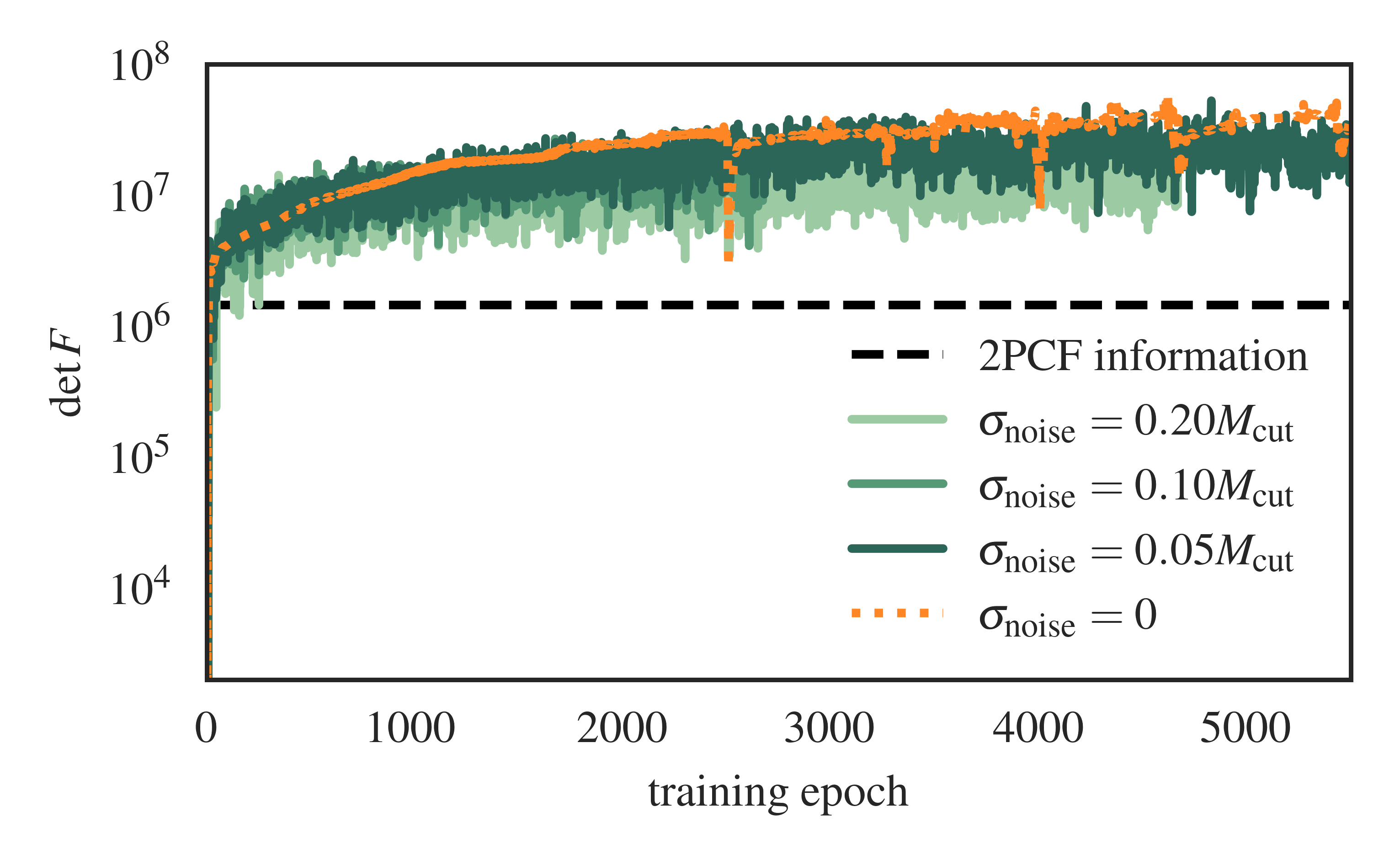}
    \caption{Validation curves for noisy masses. Smaller noise variance (darker curves) results in smaller per-epoch variance in $\det F$ and slightly more information extraction. Information leakage occurs with higher noise variance since smaller scales are poorly resolved and trimmed from the catalogue.}
    \label{fig:noisetraining}
\end{figure}

\begin{figure}
    \centering
    \includegraphics[width=\columnwidth]{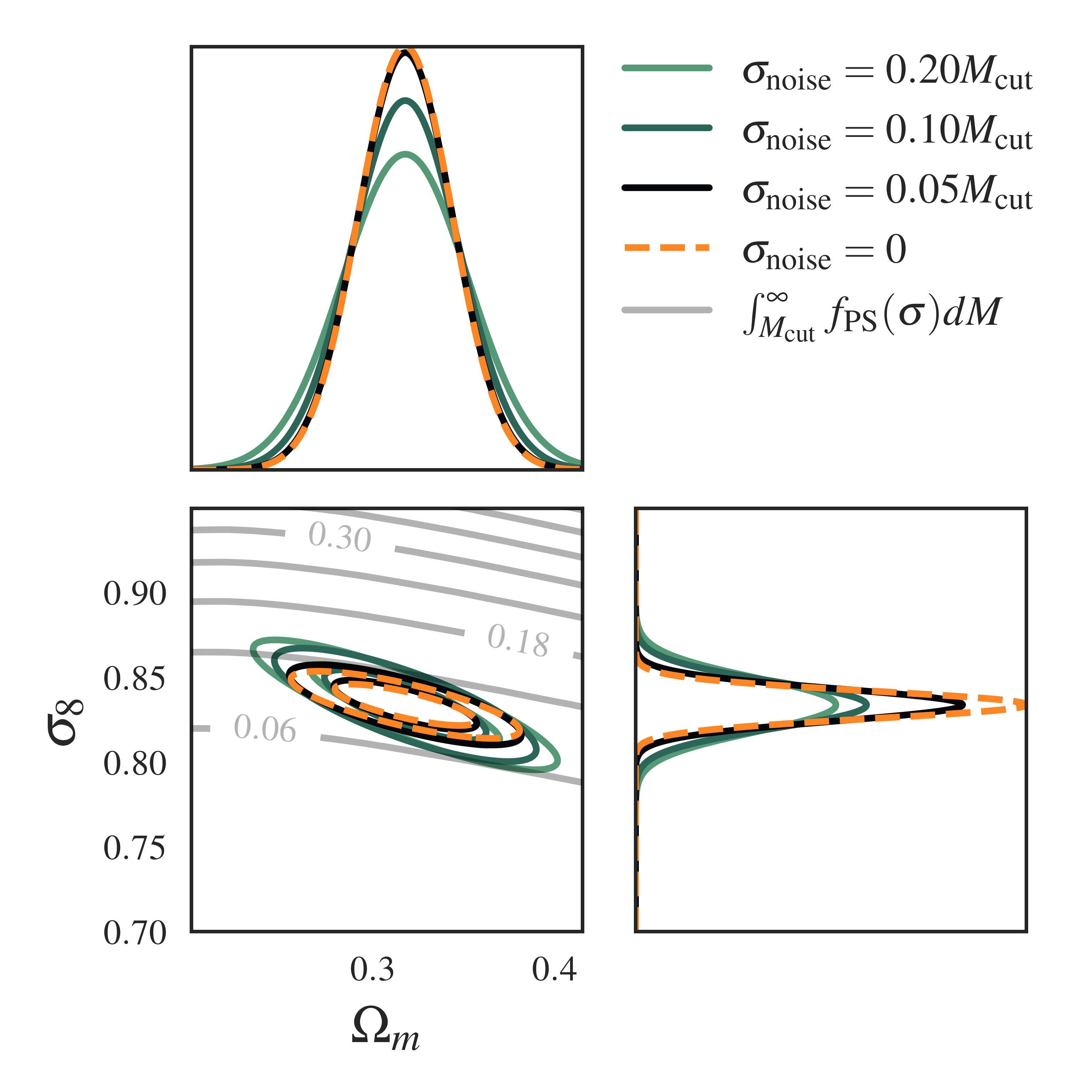}
    \caption{Fisher constraints for different noise models, plotted over lines of the Press-Schechter halo mass function integrated from $M_{\rm cut}$ (grey), where the numbers indicate the integrated mass density of halos from $M_{\rm cut}$ to infinity, as a function of $\Omega_m$ and $\sigma_8$. Higher $\sigma_{\rm noise}$ (lighter curves) cause low-mass halos to drop out, inflating constraints in $\Omega_m$, but constraints in $\sigma_8$ remain relatively unchanged since this parameter is largely position-dependent.}
    \label{fig:F_vary_mass}
\end{figure}

\noindent \textbf{Results.} We display validation curves over training epoch in Figure
\ref{fig:noisetraining}. Increasing catalogue noise results in
higher variance per epoch in the computed Fisher statistics, as
well as a slightly lower information plateau. This can be
interpreted as higher noise obscuring small mass scales in the
information extraction. This effect is illustrated via inflated
Fisher constraints in Figure \ref{fig:F_vary_mass}. 

As the noise level increases the low-end masses have more
variance when drawn on-the-fly so more halos are projected out
of the catalogue because they fall below $M_{\rm cut}$, and so
this information cannot be encapsulated in the compressed
summaries. Increasing the noise amplitude to $20\%$ of $M_{\rm cut}$ inflates constraints in $\Omega_m$. In the high-noise
limit, halo positions dominate $\det F$, indicated by the
relatively unchanged, position-dependent $\sigma_8$ constraints. As noise decreases and masses are better known,
the Fisher exhibits the same rotation seen in Section
\ref{sec:hmf-comp} along the isomass HMF lines. This effect is discussed in detail in Appendix \ref{app:hmf-comp}.

Despite inflating constraints, showing the network large
numbers of on-the-fly noise realizations during training can
harden the network to the negative effects of limited training
data and therefore provide smoother training whilst still bein
g able to extract information at a similar level to noise-free
catalogues. However, the model of these noisy masses must be
accurate to the noise model expected for the real data
otherwise the on-the-fly simulations do not provide hardening
of the summaries in the correct way and may even project out
informative data correlations.

\section{Application to Implicit Likelihood Inference}\label{sec:sbi}
\begin{figure}[htp!]
    \centering
    \includegraphics[width=\columnwidth]{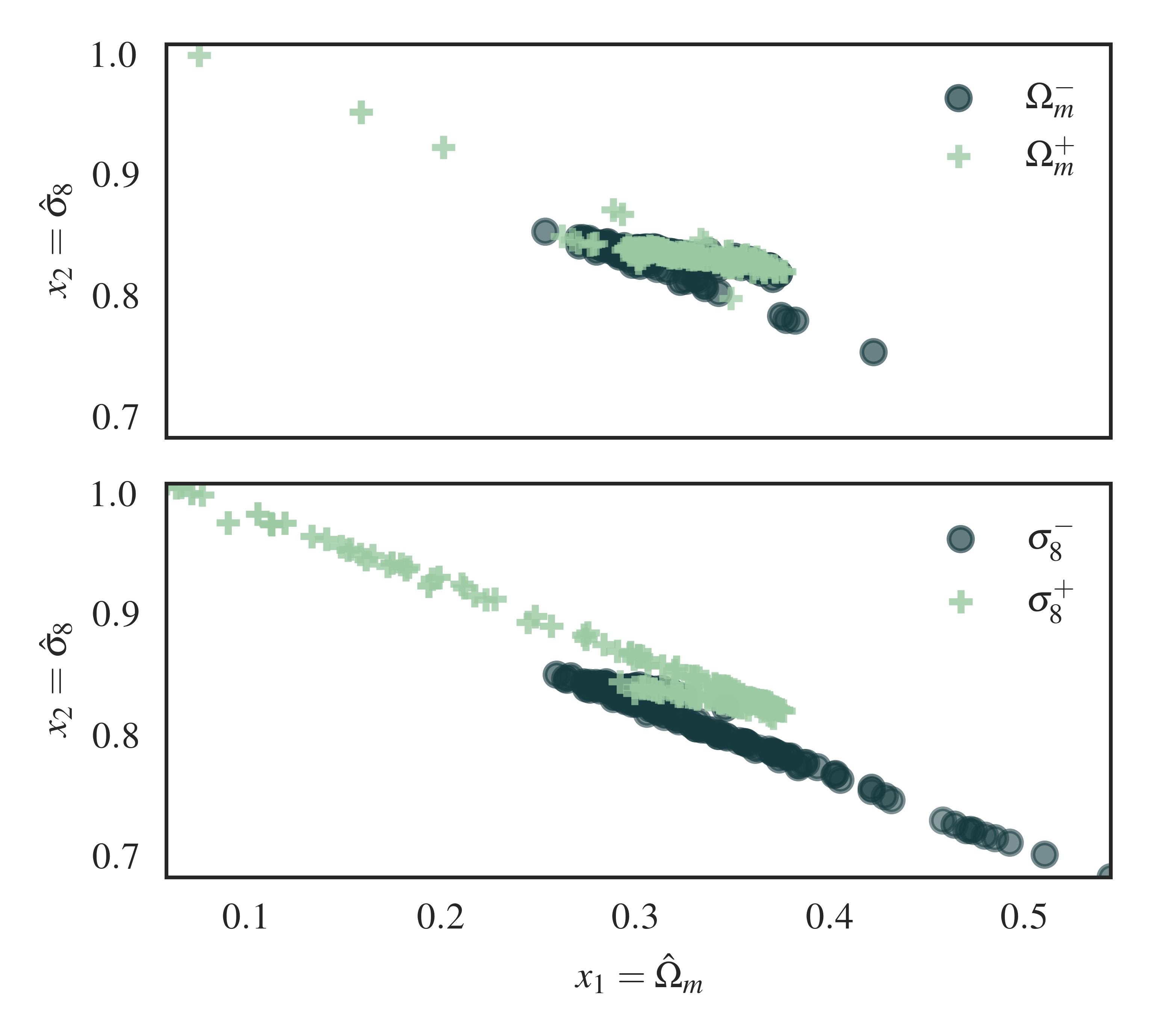}
    \caption{gIMNN score summaries for test derivative datasets ($\parvec^{\pm}$ with cosmic variance. A trained network can distinguish between simulations generated at different parameter values, indicated by the different distributions in summary space. These score summaries are \textit{not} explicit predictions for cosmological parameters; they should be considered as informative summaries of the cosmological parameters that can be used for ABC or density estimation for building cosmological posteriors. In general and on average, simulations generated at lower $\Omega_m$ and $\sigma_8$ values have lower pseudo-maximum likelihood estimated values for these parameters and vice versa, however, they should not be taken as predictions of the cosmological parameters.}
    \label{fig:summary-scatter}
\end{figure}

The IMNN framework is both an information quantification scheme as well as an {asymptotically} optimal compression mechanism for implicit likelihood inference. The global network summaries used to compute Fisher statistics can also be used as proxies for the cosmological parameters via a score estimate \citep{alsing2018_general, Charnock_IMNN} using the IMNN Fisher and covariance:
\begin{equation}
    \hat{\theta}_\alpha = \textbf{F}_{\alpha \beta}^{-1} \frac{\partial \mu_i}{\partial \theta_\beta} \textbf{C}_{ij}^{-1} \left({x}_j(\textbf{w}, \textbf{d}) - \mu_j). \right)
\end{equation}
These summaries are \textit{not} explicit predictions for cosmological parameters, although they are pseudo-maximum likelihood estimates for the parameters in the region asymptotically close to the fiducial cosmological parameter values. {Instead, we suggest using these values as informative summaries} in Approximate Bayesian Computation (ABC) or density estimation schemes, as demonstrated in \cite{makinen2021} and \cite{Charnock_IMNN}. Figure \ref{fig:summary-scatter} shows a spread of these summaries over 200 test (non-seed matched with cosmic variance) $\parvec^{\pm}$ \textit{Quijote} datasets. The network is able to distinguish halo graphs simulated at different parameter values (on the level of the parameter degeneracy), rendering these gIMNN summaries usable in accept-reject simulation-based inference (SBI) schemes. We additionally discuss network generalization to other data in Appendix \ref{app:graph-assembly}. 

\section{Discussion \& Conclusion}\label{sec:discussion}
In this study, we explored cosmological information extraction
from halo catalogues assembled as graphs. We first introduced
graphs as a general language for describing large-scale
structure formation. We showed that nonlinear summaries from
sufficiently expressive graph neural networks far exceed the
information contained in traditional 2-point statistics, using the $\approx$100 most massive halos. We illustrated that
decorating graphs with mass information increases the
information yield, as well as decreases training-validation set
sampling variance (aleatoric error). Finally, we showed that
summaries produced by the network are readily usable for
simulation-based inference.

We also explored cosmological information as a function of
graph construction. We showed that a significant amount of
information is contained in the variable cardinality of the
graph, $N^v$, {i.e. the number of halos in a catalogue,} and
related this feature to the halo mass function formalism. This
test demonstrated a distinct advantage in graph representation:
allowing data vector size to vary with cosmology,
\textit{combining both positional and mass information
automatically into just two statistics}.

Next we demonstrated that gIMNN training can be made robust to
noise with little information loss in a more realistic
setting where halo masses are estimates with measurement error.

We also explored {network (epistemic) and data sampling (aleatoric) error} in graph
representation. We showed that a combined compression of masses
and positional information \textit{decreased} data
variability, meaning training and validation graphs become more
descriptive of their underlying likelihood with decoration,
even in a fixed-length scenario. 

The results of this work hold several implications for
cosmological parameter estimation and study of large-scale
structure. The graph framework presented here enables further
modular study of nonlinear statistics that combine attributes
of mass functions and correlation functions, at a fraction of
the computational cost of bispectrum or trispectrum
calculation. Cosmological parameter constraints from void
catalogues, here the duals to halo graphs, might elucidate
complementary constraints to those obtained here
\citep{kreisch_voids}. 

\cite{quijote_png1} and \cite{quijote_png2} recently
investigated bispectrum statistics from n-body primordial
nongaussianity (pNG) simulations in the \textit{Quijote} suite
to investigate propagation of pNG to late-time cosmological
structure. A {complementary} study using gIMNNs might pick up
signatures not isolated in the bispectrum framework. The gIMNN
formalism could also be used to summarise and understand better
the halo merger and clustering statistics as a function of
scale or graph feature, such as clique number. Hierarchical
aggregation schemes, as described in Section \ref{sec:gnns},
could make the gIMNN scheme tractable for aggregating
catalogued galaxies into subhalo graphs, and further into a
global graph over the largest scales. Doing so would scale this
analysis to full-sized galaxy catalogues.

Follow-up study is warranted on much larger catalogues and with
more parameters, with a view to readying this framework for
applications to inference from detailed physical simulations
and, ultimately, realistic galaxy surveys.

\section{Code Availability}
The code used for this analysis is available at \url{https://github.com/tlmakinen/cosmicGraphs}. Full documentation for the IMNN software is available at \url{https://www.aquila-consortium.org/doc/imnn/index.html}.

\begin{acknowledgments} 
T.L.M acknowledges the Imperial College London President's Scholarship fund for support of this study, as well as the great discussions with Stephon Alexander, David Spergel, and Doug Finkbeiner that inspired this work. 
B.D.W. acknowledges support by the ANR BIG4 project, grant ANR-16-CE23-0002 of the French Agence Nationale de la Recherche;  and  the Labex ILP (reference ANR-10-LABX-63) part of the Idex SUPER, and received financial state aid managed by the Agence Nationale de la Recherche, as part of the programme Investissements d'avenir under the reference ANR-11-IDEX-0004-02. This work was done within the \href{https://aquila-consortium.org/}{Aquila Consortium} and the \href{https://www.learning-the-universe.org/}{Learning the Universe Collaboration}.
The Flatiron Institute is supported by the Simons Foundation.
\end{acknowledgments}

\bibliography{mybib}

\newcommand{\noop}[1]{}
\begin{thebibliography}{}
\expandafter\ifx\csname natexlab\endcsname\relax\def\natexlab#1{#1}\fi
\providecommand{\url}[1]{\href{#1}{#1}}
\providecommand{\dodoi}[1]{doi:~\href{http://doi.org/#1}{\nolinkurl{#1}}}
\providecommand{\doeprint}[1]{\href{http://ascl.net/#1}{\nolinkurl{http://ascl.net/#1}}}
\providecommand{\doarXiv}[1]{\href{https://arxiv.org/abs/#1}{\nolinkurl{https://arxiv.org/abs/#1}}}

\bibitem[{{Adami} \& {Mazure}(1999)}]{adami_mst_1999}
{Adami}, C., \& {Mazure}, A. 1999, Astron. Astrophys. Suppl. Ser., 134, 393,
  \dodoi{10.1051/aas:1999145}

\bibitem[{{Alpaslan} {et~al.}(2014){Alpaslan}, {Robotham}, {Driver}, {Norberg},
  {Baldry}, {Bauer}, {Bland-Hawthorn}, {Brown}, {Cluver}, {Colless}, {Foster},
  {Hopkins}, {Van Kampen}, {Kelvin}, {Lara-Lopez}, {Liske}, {Lopez-Sanchez},
  {Loveday}, {McNaught-Roberts}, {Merson}, \& {Pimbblet}}]{alpaslan_mst_2014}
{Alpaslan}, M., {Robotham}, A. S.~G., {Driver}, S., {et~al.} 2014, \mnras, 438,
  177, \dodoi{10.1093/mnras/stt2136}

\bibitem[{{Alsing} \& {Wandelt}(2018)}]{alsing2018_general}
{Alsing}, J., \& {Wandelt}, B. 2018, MNRAS, 476, L60,
  \dodoi{10.1093/mnrasl/sly029}

\bibitem[{Artis {et~al.}(2021)Artis, Melin, Bartlett, \&
  Murray}]{Artis_clusters_2021}
Artis, E., Melin, J.-B., Bartlett, J.~G., \& Murray, C. 2021, Astronomy {\&}
  Astrophysics, 649, A47, \dodoi{10.1051/0004-6361/202140293}

\bibitem[{Barrow {et~al.}(1985)Barrow, Bhavsar, \& Sonoda}]{barrow_mst_1985}
Barrow, J.~D., Bhavsar, S.~P., \& Sonoda, D.~H. 1985, Monthly Notices of the
  Royal Astronomical Society, 216, 17, \dodoi{10.1093/mnras/216.1.17}

\bibitem[{Battaglia {et~al.}(2018)Battaglia, Hamrick, Bapst, Sanchez-Gonzalez,
  Zambaldi, Malinowski, Tacchetti, Raposo, Santoro, Faulkner, Gulcehre, Song,
  Ballard, Gilmer, Dahl, Vaswani, Allen, Nash, Langston, Dyer, Heess, Wierstra,
  Kohli, Botvinick, Vinyals, Li, \& Pascanu}]{battaglia2018relational}
Battaglia, P.~W., Hamrick, J.~B., Bapst, V., {et~al.} 2018, Relational
  inductive biases, deep learning, and graph networks.
\newblock \doarXiv{1806.01261}

\bibitem[{{Beuret} {et~al.}(2017){Beuret}, {Billot}, {Cambr{\'e}sy}, {Eden},
  {Elia}, {Molinari}, {Pezzuto}, \& {Schisano}}]{beuret_mst_2017}
{Beuret}, M., {Billot}, N., {Cambr{\'e}sy}, L., {et~al.} 2017, \aap, 597, A114,
  \dodoi{10.1051/0004-6361/201629199}

\bibitem[{{Bhavsar} \& {Ling}(1988)}]{bhavsar_1988mst}
{Bhavsar}, S.~P., \& {Ling}, E.~N. 1988, \pasp, 100, 1314,
  \dodoi{10.1086/132325}

\bibitem[{{Biswas} {et~al.}(2010){Biswas}, {Alizadeh}, \&
  {Wandelt}}]{2010PhRvD..82b3002B}
{Biswas}, R., {Alizadeh}, E., \& {Wandelt}, B.~D. 2010, \prd, 82, 023002,
  \dodoi{10.1103/PhysRevD.82.023002}

\bibitem[{Bonnaire {et~al.}(2020)Bonnaire, Aghanim, Decelle, \&
  Douspis}]{Bonnaire_2020}
Bonnaire, T., Aghanim, N., Decelle, A., \& Douspis, M. 2020, Astronomy {\&}
  Astrophysics, 637, A18, \dodoi{10.1051/0004-6361/201936859}

\bibitem[{Bonnaire {et~al.}(2022)Bonnaire, Aghanim, Kuruvilla, \&
  Decelle}]{Bonnaire_2022}
Bonnaire, T., Aghanim, N., Kuruvilla, J., \& Decelle, A. 2022, Astronomy {\&}
  Astrophysics, 661, A146, \dodoi{10.1051/0004-6361/202142852}

\bibitem[{Bronstein {et~al.}(2021)Bronstein, Bruna, Cohen, \&
  Veličković}]{bronstein2021_permeq}
Bronstein, M.~M., Bruna, J., Cohen, T., \& Veličković, P. 2021, Geometric
  Deep Learning: Grids, Groups, Graphs, Geodesics, and Gauges,  arXiv,
  \dodoi{10.48550/ARXIV.2104.13478}

\bibitem[{Charnock {et~al.}(2018)Charnock, Lavaux, \& Wandelt}]{Charnock_IMNN}
Charnock, T., Lavaux, G., \& Wandelt, B.~D. 2018, Physical Review D, 97,
  \dodoi{10.1103/physrevd.97.083004}

\bibitem[{Colberg(2007)}]{colberg_mst_2007}
Colberg, J.~M. 2007, Monthly Notices of the Royal Astronomical Society, 375,
  337, \dodoi{10.1111/j.1365-2966.2006.11312.x}

\bibitem[{{Coles} {et~al.}(1998){Coles}, {Pearson}, {Borgani}, {Plionis}, \&
  {Moscardini}}]{coles_1998_mst}
{Coles}, P., {Pearson}, R.~C., {Borgani}, S., {Plionis}, M., \& {Moscardini},
  L. 1998, \mnras, 294, 245, \dodoi{10.1046/j.1365-8711.1998.01147.x}

\bibitem[{Coulton {et~al.}(2022)Coulton, Villaescusa-Navarro, Jamieson, Baldi,
  Jung, Karagiannis, Liguori, Verde, \& Wandelt}]{quijote_png1}
Coulton, W.~R., Villaescusa-Navarro, F., Jamieson, D., {et~al.} 2022,
  Quijote-PNG: Simulations of primordial non-Gaussianity and the information
  content of the matter field power spectrum and bispectrum,  arXiv,
  \dodoi{10.48550/ARXIV.2206.01619}

\bibitem[{Cramér(1946)}]{cramerharald_1946}
Cramér, H. 1946, Mathematical methods of statistics, by Harald Cramer, .. (The
  University Press)

\bibitem[{Cranmer {et~al.}(2020)Cranmer, Sanchez-Gonzalez, Battaglia, Xu,
  Cranmer, Spergel, \& Ho}]{cranmer2020discovering}
Cranmer, M., Sanchez-Gonzalez, A., Battaglia, P., {et~al.} 2020, Discovering
  Symbolic Models from Deep Learning with Inductive Biases.
\newblock \doarXiv{2006.11287}

\bibitem[{{Dai} \& {Seljak}(2022)}]{seljak_trenf}
{Dai}, B., \& {Seljak}, U. 2022, arXiv e-prints, arXiv:2202.05282.
\newblock \doarXiv{2202.05282}

\bibitem[{{Davis} {et~al.}(1985){Davis}, {Efstathiou}, {Frenk}, \&
  {White}}]{davis_fof}
{Davis}, M., {Efstathiou}, G., {Frenk}, C.~S., \& {White}, S.~D.~M. 1985, \apj,
  292, 371, \dodoi{10.1086/163168}

\bibitem[{Fluri {et~al.}(2019)Fluri, Kacprzak, Lucchi, Refregier, Amara,
  Hofmann, \& Schneider}]{Fluri_2019lensing}
Fluri, J., Kacprzak, T., Lucchi, A., {et~al.} 2019, Physical Review D, 100,
  \dodoi{10.1103/physrevd.100.063514}

\bibitem[{{Fluri} {et~al.}(2022){Fluri}, {Kacprzak}, {Lucchi}, {Schneider},
  {Refregier}, \& {Hofmann}}]{fluri_kids}
{Fluri}, J., {Kacprzak}, T., {Lucchi}, A., {et~al.} 2022, arXiv e-prints,
  arXiv:2201.07771.
\newblock \doarXiv{2201.07771}

\bibitem[{Fluri {et~al.}(2018)Fluri, Kacprzak, Refregier, Amara, Lucchi, \&
  Hofmann}]{Fluri_2018}
Fluri, J., Kacprzak, T., Refregier, A., {et~al.} 2018, Physical Review D, 98,
  \dodoi{10.1103/physrevd.98.123518}

\bibitem[{Fluri {et~al.}(2021)Fluri, Kacprzak, Refregier, Lucchi, \&
  Hofmann}]{fluri2021}
Fluri, J., Kacprzak, T., Refregier, A., Lucchi, A., \& Hofmann, T. 2021,
  Physical Review D, 104, \dodoi{10.1103/physrevd.104.123526}

\bibitem[{Gillet {et~al.}(2019)Gillet, Mesinger, Greig, Liu, \&
  Ucci}]{Gillet_2019}
Gillet, N., Mesinger, A., Greig, B., Liu, A., \& Ucci, G. 2019, Monthly Notices
  of the Royal Astronomical Society, \dodoi{10.1093/mnras/stz010}

\bibitem[{Godwin {et~al.}(2020)Godwin, Keck, Battaglia, Bapst, Kipf, Li,
  Stachenfeld, Velikovic, \& Sanchez-Gonzalez}]{jraph2020github}
Godwin, J., Keck, T., Battaglia, P., {et~al.} 2020, {J}raph: {A} library for
  graph neural networks in jax., 0.0.1.dev.
\newblock \url{http://github.com/deepmind/jraph}

\bibitem[{Hahn {et~al.}(2020)Hahn, Villaescusa-Navarro, Castorina, \&
  Scoccimarro}]{bispecHahn_2020}
Hahn, C., Villaescusa-Navarro, F., Castorina, E., \& Scoccimarro, R. 2020,
  Journal of Cosmology and Astroparticle Physics, 2020, 040,
  \dodoi{10.1088/1475-7516/2020/03/040}

\bibitem[{Hamaus {et~al.}(2015)Hamaus, Sutter, Lavaux, \&
  Wandelt}]{hamaus_voids2015}
Hamaus, N., Sutter, P., Lavaux, G., \& Wandelt, B.~D. 2015, Journal of
  Cosmology and Astroparticle Physics, 2015, 036–036,
  \dodoi{10.1088/1475-7516/2015/11/036}

\bibitem[{Heavens {et~al.}(2000)Heavens, Jimenez, \& Lahav}]{Heavens_2000}
Heavens, A.~F., Jimenez, R., \& Lahav, O. 2000, Monthly Notices of the Royal
  Astronomical Society, 317, 965–972,
  \dodoi{10.1046/j.1365-8711.2000.03692.x}

\bibitem[{Heek {et~al.}(2020)Heek, Levskaya, Oliver, Ritter, Rondepierre,
  Steiner, \& van {Z}ee}]{flax2020github}
Heek, J., Levskaya, A., Oliver, A., {et~al.} 2020, {F}lax: A neural network
  library and ecosystem for {JAX}, 0.5.2.
\newblock \url{http://github.com/google/flax}

\bibitem[{Hendrycks \& Gimpel(2016)}]{gelu_ref}
Hendrycks, D., \& Gimpel, K. 2016, Gaussian Error Linear Units (GELUs),  arXiv,
  \dodoi{10.48550/ARXIV.1606.08415}

\bibitem[{Jamieson {et~al.}(2022)Jamieson, Li, de~Oliveira,
  Villaescusa-Navarro, Ho, \& Spergel}]{shirleyneural_emu}
Jamieson, D., Li, Y., de~Oliveira, R.~A., {et~al.} 2022, Field Level Neural
  Network Emulator for Cosmological N-body Simulations,  arXiv,
  \dodoi{10.48550/ARXIV.2206.04594}

\bibitem[{Jasche {et~al.}(2015)Jasche, Leclercq, \& Wandelt}]{Jasche_2015}
Jasche, J., Leclercq, F., \& Wandelt, B. 2015, Journal of Cosmology and
  Astroparticle Physics, 2015, 036, \dodoi{10.1088/1475-7516/2015/01/036}

\bibitem[{{Jasche} \& {Wandelt}(2013)}]{2013MNRAS.432..894J}
{Jasche}, J., \& {Wandelt}, B.~D. 2013, \mnras, 432, 894,
  \dodoi{10.1093/mnras/stt449}

\bibitem[{Jeffrey {et~al.}(2020)Jeffrey, Alsing, \&
  Lanusse}]{Jeffrey_2020LFI_field}
Jeffrey, N., Alsing, J., \& Lanusse, F. 2020, Monthly Notices of the Royal
  Astronomical Society, 501, 954, \dodoi{10.1093/mnras/staa3594}

\bibitem[{{Jeffrey} {et~al.}(2022){Jeffrey}, {Boulanger}, {Wandelt},
  {Regaldo-Saint Blancard}, {Allys}, \& {Levrier}}]{2022MNRAS.510L...1J}
{Jeffrey}, N., {Boulanger}, F., {Wandelt}, B.~D., {et~al.} 2022, \mnras, 510,
  L1, \dodoi{10.1093/mnrasl/slab120}

\bibitem[{Jeffrey \& Wandelt(2020)}]{jeffrey2020solving}
Jeffrey, N., \& Wandelt, B.~D. 2020, Solving high-dimensional parameter
  inference: marginal posterior densities \& Moment Networks.
\newblock \doarXiv{2011.05991}

\bibitem[{Jung {et~al.}(2022)Jung, Karagiannis, Liguori, Baldi, Coulton,
  Jamieson, Verde, Villaescusa-Navarro, \& Wandelt}]{quijote_png2}
Jung, G., Karagiannis, D., Liguori, M., {et~al.} 2022, Quijote-PNG:
  Quasi-maximum likelihood estimation of Primordial Non-Gaussianity in the
  non-linear dark matter density field,  arXiv,
  \dodoi{10.48550/ARXIV.2206.01624}

\bibitem[{Kipf \& Welling(2016)}]{kipf_gcn}
Kipf, T.~N., \& Welling, M. 2016, Semi-Supervised Classification with Graph
  Convolutional Networks,  arXiv, \dodoi{10.48550/ARXIV.1609.02907}

\bibitem[{Kreisch {et~al.}(2021)Kreisch, Pisani, Villaescusa-Navarro, Spergel,
  Wandelt, Hamaus, \& Bayer}]{kreisch_voids}
Kreisch, C.~D., Pisani, A., Villaescusa-Navarro, F., {et~al.} 2021, The
  GIGANTES dataset: precision cosmology from voids in the machine learning era,
   arXiv, \dodoi{10.48550/ARXIV.2107.02304}

\bibitem[{{Krzewina} \& {Saslaw}(1996)}]{kyrzewena_mst}
{Krzewina}, L.~G., \& {Saslaw}, W.~C. 1996, \mnras, 278, 869,
  \dodoi{10.1093/mnras/278.3.869}

\bibitem[{Kwon {et~al.}(2020)Kwon, Hong, \& Park}]{Kwon_2020}
Kwon, Y., Hong, S.~E., \& Park, I. 2020, Journal of the Korean Physical
  Society, 77, 49–59, \dodoi{10.3938/jkps.77.49}

\bibitem[{{Lavaux} \& {Wandelt}(2010)}]{2010MNRAS.403.1392L}
{Lavaux}, G., \& {Wandelt}, B.~D. 2010, \mnras, 403, 1392,
  \dodoi{10.1111/j.1365-2966.2010.16197.x}

\bibitem[{Leclercq(2015)}]{florent_borg}
Leclercq, F. 2015, Bayesian large-scale structure inference and cosmic web
  analysis,  arXiv, \dodoi{10.48550/ARXIV.1512.04985}

\bibitem[{Leclercq \& Heavens(2021)}]{leclercq2021accuracy}
Leclercq, F., \& Heavens, A. 2021, On the accuracy and precision of correlation
  functions and field-level inference in cosmology.
\newblock \doarXiv{2103.04158}

\bibitem[{Lemos {et~al.}(2022)Lemos, Jeffrey, Cranmer, Ho, \&
  Battaglia}]{lemos2022rediscovering}
Lemos, P., Jeffrey, N., Cranmer, M., Ho, S., \& Battaglia, P. 2022,
  Rediscovering orbital mechanics with machine learning.
\newblock \doarXiv{2202.02306}

\bibitem[{{Libeskind} {et~al.}(2018){Libeskind}, {van de Weygaert}, {Cautun},
  {Falck}, {Tempel}, {Abel}, {Alpaslan}, {Arag{\'o}n-Calvo}, {Forero-Romero},
  {Gonzalez}, {Gottl{\"o}ber}, {Hahn}, {Hellwing}, {Hoffman}, {Jones},
  {Kitaura}, {Knebe}, {Manti}, {Neyrinck}, {Nuza}, {Padilla}, {Platen},
  {Ramachandra}, {Robotham}, {Saar}, {Shandarin}, {Steinmetz}, {Stoica},
  {Sousbie}, \& {Yepes}}]{libeskind_web}
{Libeskind}, N.~I., {van de Weygaert}, R., {Cautun}, M., {et~al.} 2018, \mnras,
  473, 1195, \dodoi{10.1093/mnras/stx1976}

\bibitem[{Livet {et~al.}(2021)Livet, Charnock, Borgne, \&
  de~Lapparent}]{livet2021catalogfree}
Livet, F., Charnock, T., Borgne, D.~L., \& de~Lapparent, V. 2021, Catalog-free
  modeling of galaxy types in deep images: Massive dimensional reduction with
  neural networks.
\newblock \doarXiv{2102.01086}

\bibitem[{Makinen {et~al.}(2021)Makinen, Charnock, Alsing, \&
  Wandelt}]{makinen2021}
Makinen, T.~L., Charnock, T., Alsing, J., \& Wandelt, B.~D. 2021, Journal of
  Cosmology and Astroparticle Physics, 2021, 049,
  \dodoi{10.1088/1475-7516/2021/11/049}

\bibitem[{Makinen {et~al.}(2020)Makinen, Lancaster, Villaescusa-Navarro,
  Melchior, Ho, Perreault-Levasseur, \& Spergel}]{makinen2020deep21}
Makinen, T.~L., Lancaster, L., Villaescusa-Navarro, F., {et~al.} 2020, deep21:
  a Deep Learning Method for 21cm Foreground Removal.
\newblock \doarXiv{2010.15843}

\bibitem[{Massara {et~al.}(2022)Massara, Villaescusa-Navarro, Hahn, Abidi,
  Eickenberg, Ho, Lemos, Dizgah, \& Blancard}]{marked_pk_quijote}
Massara, E., Villaescusa-Navarro, F., Hahn, C., {et~al.} 2022, Cosmological
  Information in the Marked Power Spectrum of the Galaxy Field,  arXiv,
  \dodoi{10.48550/ARXIV.2206.01709}

\bibitem[{Matilla {et~al.}(2020)Matilla, Sharma, Hsu, \& Haiman}]{Matilla_2020}
Matilla, J. M.~Z., Sharma, M., Hsu, D., \& Haiman, Z. 2020, Physical Review D,
  102, \dodoi{10.1103/physrevd.102.123506}

\bibitem[{{Murray}(2014)}]{hmfcalc_soft}
{Murray}, S. 2014, {HMF: Halo Mass Function calculator}, Astrophysics Source
  Code Library, record ascl:1412.006.
\newblock \doeprint{1412.006}

\bibitem[{{Murray} {et~al.}(2013){Murray}, {Power}, \&
  {Robotham}}]{hmf_calc_murray}
{Murray}, S.~G., {Power}, C., \& {Robotham}, A.~S.~G. 2013, Astronomy and
  Computing, 3, 23, \dodoi{10.1016/j.ascom.2013.11.001}

\bibitem[{Naidoo {et~al.}(2022)Naidoo, Massara, \&
  Lahav}]{cosmo_from_graphsNaidoo_2022}
Naidoo, K., Massara, E., \& Lahav, O. 2022, Monthly Notices of the Royal
  Astronomical Society, 513, 3596, \dodoi{10.1093/mnras/stac1138}

\bibitem[{Naidoo {et~al.}(2019)Naidoo, Whiteway, Massara, Gualdi, Lahav, Viel,
  Gil-Marín, \& Font-Ribera}]{naidoo_mst_2020}
Naidoo, K., Whiteway, L., Massara, E., {et~al.} 2019, Monthly Notices of the
  Royal Astronomical Society, 491, 1709, \dodoi{10.1093/mnras/stz3075}

\bibitem[{Pan {et~al.}(2020)Pan, Liu, Forero-Romero, Sabiu, Li, Miao, \&
  Li}]{pan2020cosmological}
Pan, S., Liu, M., Forero-Romero, J., {et~al.} 2020, Cosmological parameter
  estimation from large-scale structure deep learning.
\newblock \doarXiv{1908.10590}

\bibitem[{Petri {et~al.}(2013)Petri, Haiman, Hui, May, \&
  Kratochvil}]{petri_minkowski}
Petri, A., Haiman, Z., Hui, L., May, M., \& Kratochvil, J.~M. 2013, Phys. Rev.
  D, 88, 123002, \dodoi{10.1103/PhysRevD.88.123002}

\bibitem[{Philcox \& Ivanov(2022)}]{philcox_bispectra2022}
Philcox, O.~H., \& Ivanov, M.~M. 2022, Physical Review D, 105,
  \dodoi{10.1103/physrevd.105.043517}

\bibitem[{Porqueres {et~al.}(2021)Porqueres, Heavens, Mortlock, \&
  Lavaux}]{natalia2021}
Porqueres, N., Heavens, A., Mortlock, D., \& Lavaux, G. 2021, Monthly Notices
  of the Royal Astronomical Society, 509, 3194–3202,
  \dodoi{10.1093/mnras/stab3234}

\bibitem[{Prelogović {et~al.}(2021)Prelogović, Mesinger, Murray, Fiameni, \&
  Gillet}]{prelogovic2021machine}
Prelogović, D., Mesinger, A., Murray, S., Fiameni, G., \& Gillet, N. 2021,
  Machine learning galaxy properties from 21 cm lightcones: impact of network
  architectures and signal contamination.
\newblock \doarXiv{2107.00018}

\bibitem[{{Press} \& {Schechter}(1974)}]{press_schechter_fn}
{Press}, W.~H., \& {Schechter}, P. 1974, \apj, 187, 425, \dodoi{10.1086/152650}

\bibitem[{{Ramanah} {et~al.}(2019){Ramanah}, {Lavaux}, {Jasche}, \&
  {Wandelt}}]{2019A&A...621A..69R}
{Ramanah}, D.~K., {Lavaux}, G., {Jasche}, J., \& {Wandelt}, B.~D. 2019, \aap,
  621, A69, \dodoi{10.1051/0004-6361/201834117}

\bibitem[{Rao(1945)}]{rao_1945}
Rao, C.~R. 1945, Bulletin of the Calcutta Mathematical Society, 37, 81–89

\bibitem[{Ravanbakhsh {et~al.}(2017)Ravanbakhsh, Oliva, Fromenteau, Price, Ho,
  Schneider, \& Poczos}]{ravanbakhsh2017estimating}
Ravanbakhsh, S., Oliva, J., Fromenteau, S., {et~al.} 2017, Estimating
  Cosmological Parameters from the Dark Matter Distribution.
\newblock \doarXiv{1711.02033}

\bibitem[{Reed {et~al.}(2006)Reed, Bower, Frenk, Jenkins, \&
  Theuns}]{halomass_cosmo}
Reed, D.~S., Bower, R., Frenk, C.~S., Jenkins, A., \& Theuns, T. 2006, Monthly
  Notices of the Royal Astronomical Society, 374, 2,
  \dodoi{10.1111/j.1365-2966.2006.11204.x}

\bibitem[{Ribli {et~al.}(2018)Ribli, Ármin Pataki, \&
  Csabai}]{ribli2018improved}
Ribli, D., Ármin Pataki, B., \& Csabai, I. 2018, An improved cosmological
  parameter inference scheme motivated by deep learning.
\newblock \doarXiv{1806.05995}

\bibitem[{Satorras {et~al.}(2021)Satorras, Hoogeboom, Fuchs, Posner, \&
  Welling}]{argmax_flow}
Satorras, V.~G., Hoogeboom, E., Fuchs, F.~B., Posner, I., \& Welling, M. 2021,
  E(n) Equivariant Normalizing Flows,  arXiv, \dodoi{10.48550/ARXIV.2105.09016}

\bibitem[{{Sutter} {et~al.}(2012){Sutter}, {Lavaux}, {Wandelt}, \&
  {Weinberg}}]{2012ApJ...761...44S}
{Sutter}, P.~M., {Lavaux}, G., {Wandelt}, B.~D., \& {Weinberg}, D.~H. 2012,
  \apj, 761, 44, \dodoi{10.1088/0004-637X/761/1/44}

\bibitem[{Tegmark {et~al.}(1997)Tegmark, Taylor, \& Heavens}]{Tegmark_1997}
Tegmark, M., Taylor, A.~N., \& Heavens, A.~F. 1997, The Astrophysical Journal,
  480, 22–35, \dodoi{10.1086/303939}

\bibitem[{Ueda \& Itoh(1997)}]{itoh_mst}
Ueda, H., \& Itoh, M. 1997, Publications of the Astronomical Society of Japan,
  49, 131, \dodoi{10.1093/pasj/49.2.131}

\bibitem[{{Uhlemann} {et~al.}(2020){Uhlemann}, {Friedrich},
  {Villaescusa-Navarro}, {Banerjee}, \& {Codis}}]{uhlemann_1dprob}
{Uhlemann}, C., {Friedrich}, O., {Villaescusa-Navarro}, F., {Banerjee}, A., \&
  {Codis}, S. 2020, \mnras, 495, 4006, \dodoi{10.1093/mnras/staa1155}

\bibitem[{Uhlemann {et~al.}(2020)Uhlemann, Friedrich, Villaescusa-Navarro,
  Banerjee, \& Codis}]{countsincells_fisher}
Uhlemann, C., Friedrich, O., Villaescusa-Navarro, F., Banerjee, A., \& Codis,
  S. 2020, Monthly Notices of the Royal Astronomical Society, 495, 4006,
  \dodoi{10.1093/mnras/staa1155}

\bibitem[{{van de Weygaert} {et~al.}(1992){van de Weygaert}, Jones, \&
  Martínez}]{vandeweygaet_mst}
{van de Weygaert}, R., Jones, B.~J., \& Martínez, V.~J. 1992, Physics Letters
  A, 169, 145, \dodoi{https://doi.org/10.1016/0375-9601(92)90584-9}

\bibitem[{Villaescusa-Navarro {et~al.}(2020{\natexlab{a}})Villaescusa-Navarro,
  Wandelt, Anglés-Alcázar, Genel, Mantilla, Ho, \&
  Spergel}]{villaescusanavarro2020neural}
Villaescusa-Navarro, F., Wandelt, B.~D., Anglés-Alcázar, D., {et~al.}
  2020{\natexlab{a}}, Neural networks as optimal estimators to marginalize over
  baryonic effects.
\newblock \doarXiv{2011.05992}

\bibitem[{Villaescusa-Navarro {et~al.}(2020{\natexlab{b}})Villaescusa-Navarro,
  Hahn, Massara, Banerjee, Delgado, Ramanah, Charnock, Giusarma, Li, Allys,
  Brochard, Uhlemann, Chiang, He, Pisani, Obuljen, Feng, Castorina, Contardo,
  Kreisch, Nicola, Alsing, Scoccimarro, Verde, Viel, Ho, Mallat, Wandelt, \&
  Spergel}]{quijote2020}
Villaescusa-Navarro, F., Hahn, C., Massara, E., {et~al.} 2020{\natexlab{b}},
  The Astrophysical Journal Supplement Series, 250, 2,
  \dodoi{10.3847/1538-4365/ab9d82}

\bibitem[{Villanueva-Domingo \& Villaescusa-Navarro(2022)}]{paco_graphs}
Villanueva-Domingo, P., \& Villaescusa-Navarro, F. 2022, Learning cosmology and
  clustering with cosmic graphs,  arXiv, \dodoi{10.48550/ARXIV.2204.13713}

\bibitem[{Xu {et~al.}(2019)Xu, Cisewski-Kehe, Green, \&
  Nagai}]{findvoids_topology2019}
Xu, X., Cisewski-Kehe, J., Green, S., \& Nagai, D. 2019, Astronomy and
  Computing, 27, 34–52, \dodoi{10.1016/j.ascom.2019.02.003}

\bibitem[{Yang \& Yu(2022)}]{graph_for_halos22}
Yang, D., \& Yu, H.-B. 2022, A graph model for the clustering of dark matter
  halos,  arXiv, \dodoi{10.48550/ARXIV.2206.05578}

\end{thebibliography}
\bibliographystyle{aasjournal}

\myemptypage
\newpage
\appendix

\begin{figure}[htp!]
    \centering
    \includegraphics[width=\columnwidth]{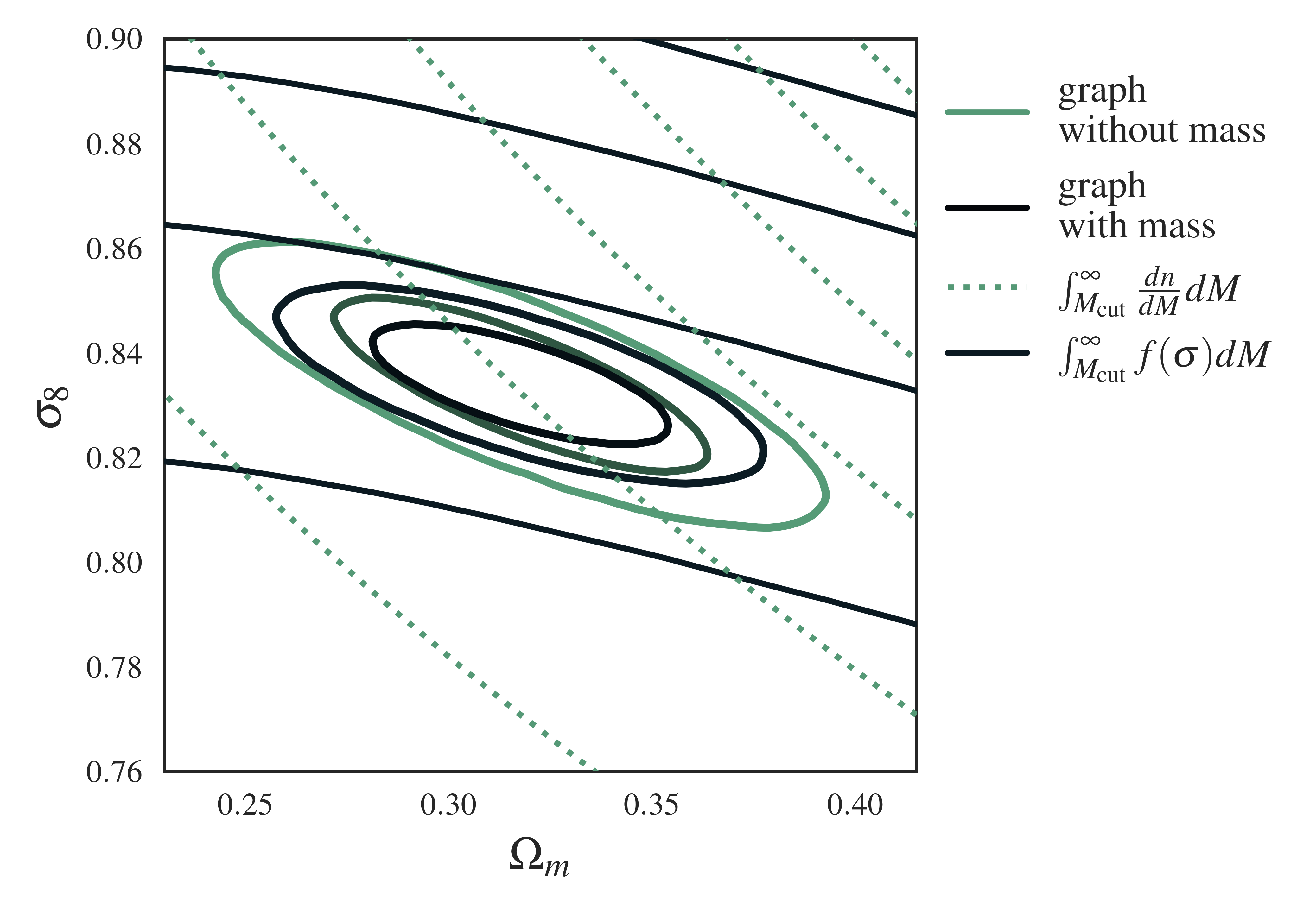}
    \caption{Comparison of undecorated graph (green ellipse) and decorated graph (dark ellipse) gIMNN
    Fishers, plotted over integrated halo number density
    (dashed green lines) and mass fraction (dark solid lines) functions from fixed $M_{\rm cut}$. Decorating
    graphs with mass information induces a slight
    rotation away from $dn/dM$ and towards $f(\sigma)$
    contours, since this quantity incorporates both mass
    and halo number information.}
    \label{fig:fsigma-dndm-comp}
\end{figure}
\section{Comparing Halo Mass and Number Density Functions}\label{app:hmf-comp}
Here we discuss the difference between the halo number density function and halo mass function in the context of gIMNN information extraction. The halo number density function, 
\begin{equation}\label{eq:dndm}
    \frac{dn}{dM} = \frac{\rho_o}{M}\frac{d\ln \sigma^{-1}(M)}{dM} f(\sigma),
\end{equation}
and the halo mass function, 
\begin{equation}
    f(\sigma) = \frac{M}{\rho_o} \frac{dn(M)}{d \ln (\sigma^{-1}(M))},
\end{equation}
within the Press-Schechter formalism \citep{press_schechter_fn}.
Integrating the halo number density from a fixed mass $M_{\rm cut}$ yields the number of halos with a mass above this threshold:
\begin{equation}
    N(M_i > M_{\rm cut}) = \int_{M_{\rm cut}}^\infty \frac{dn}{dM} dM,
\end{equation}
which in our case is the node cardinality of a halo graph, $N^v$.
By contrast, integrating the halo mass function from $M_{\rm cut}$ yields the \textit{fraction of total mass} residing in collapsed halos of mass above $M_{\rm cut}$:
\begin{equation}
    F(M_i > M_{\rm cut}) = \int_{M_{\rm cut}}^\infty f(\sigma) dM,
\end{equation}
which incorporates \textit{both} halo number $N^v$ and mass information above $M_{\rm cut}$.

We compare these two integrated quantities to undecorated and decorated gIMNN Fisher constraints in Figure \ref{fig:fsigma-dndm-comp}. In the undecorated case
(green ellipse), the network has explicit access to halo number
and clustering information, resulting in contours more
closely aligned with integrated $dn/dM$ (dashed green
lines). By contrast, when the graph nodes are annotated
with mass labels (black ellipse), the network has explicit access
to a combination of halo number, mass, and clustering
information. This results in a slight rotation towards
the integrated HMF $f(\sigma)$ contours (dark solid
lines), since this quantity reflects the addition of mass
fraction information. This effect is also illustrated in
Figure \ref{fig:F_vary_mass}. As discussed in Section
\ref{sec:withnoise}, as noise level decreases, the
network has access to sharper mass information, inducing
a rotation towards the integrated HMF line. 

\section{Details of Graph Assembly in \texttt{Jax}}\label{app:graph-assembly}
Here we detail \texttt{Jax}-compatible graph assembly.
\texttt{Jax} is a pseudo-compiled language, meaning arrays
must have a pre-determined fixed length before sent to a
GPU device for operations like gradient descent. We
navigate this constraint by padding graph features by
pre-determined fixed values, and masking features to
assess information content in a modular fashion.

\begin{figure}[htp!]
    \centering
    \includegraphics[width=\columnwidth]{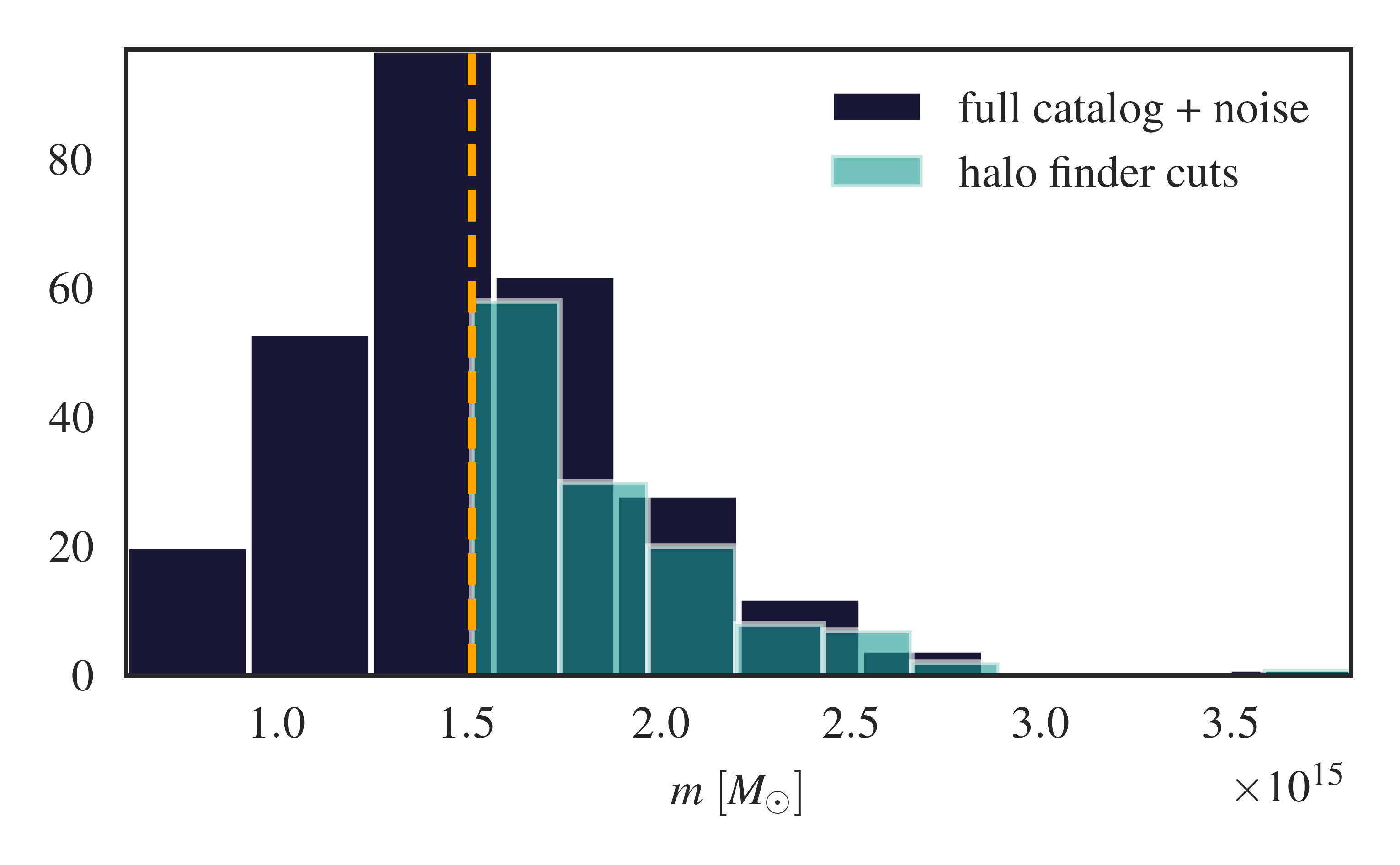}
    \caption{Mass distribution after added noise, $\sigma_{\rm noise}=0.2 M_{\rm cut}$ (black) and simulated halo finder cuts (teal) for a single fiducial simulation for masses larger than $1.1\times 10^{15}\ M_\odot$. The orange dashed lines indicates the minimum mass considered by the ``survey'' cutoff, $M_{\rm cut}=1.5\times 10^{15}\ M_\odot$.}
    \label{fig:masscut}
\end{figure}

The \textit{Quijote} catalogues are assembled into graphs by
first making a mass cut on a larger selection of halos.
Masses are then padded with a pre-determined padding
value, generally chosen to be $\max(N^v) + 10$. These
dummy halos are assigned a very large mass value and a
position outside of the $1 \rm Gpc^3$ box. A distance
matrix is then computed for \textit{both} halo and dummy
halo nodes, along with sender-receiver indexes, ($s_k,r_k$). In the invariant graph case we also compute relative angles between all halos, outlined in Section \ref{sec:edgeconstruct}. Connections $|\textbf{d}_{ij}| > r_{\rm connect}$ and $\textbf{d}_{ij}=0$ (self-edges) are then
removed and replaced by a large padding value larger than $r_{\rm connect}$. We compute $N^e$ by the number of distances that fit these criteria. Distance values and sender-receiver arrays are then sorted smallest to largest and slotted into padded arrays of length $\max(N^e) + 10$, where padded edge values are then replaced with zeros. We encountered gradient stability issues when padding was made too large for e.g. smaller edge sets, since more padding means the network is asked to operate on extra non-informative features. This sorting arrangement is advantageous since small distances (local connections) are always included in the halo graph's edges, even if the padding container is chosen to be too small for all edges for a given $r_{\rm connect}$ value. What this means is that networks trained on a small $r_{\rm connect}$ can generalize reasonably well to datasets constructed with larger connection criteria.

\noindent \textbf{Adding Noise.} When constructing noisy catalogues on-the-fly, we first truncate catalogues with a smaller $M_{\rm cut}=1.1\times 10^{15}M_\odot$ to include more halos in the true catalog. Every training epoch, a new realisation of observational noise is then added to the masses according to Section \ref{sec:withnoise}, after which halos below $M_{\rm cut}=1.5\times 10^{15}M_\odot$ are discarded, illustrated in Figure \ref{fig:masscut}. The graph edge attributes are then computed for the remaining halos as described above. For the noise scheme chosen in this work, this results in approximately equal numbers of halos being discarded above and below the mass cut line, yet increases the uncertainty in the informative HMF number count, which inflates constraints in $\Omega_m$.

\section{Details of the GNN Architecture}\label{app:gnn-details}
\noindent \textbf{Masking Graph Features.} To conduct information extraction tests with and without node decoration, the GNN architecture must remain fixed between decorated and undecorated cases. To mask a graph node or edge feature, an indicator $\textbf{1}$ is assigned to the array in place of numerical value. These uninformative features are fed through the network but do not contribute to the information extraction since the same operation is performed across fiducial and derivative datasets, yet ensure a fair comparison of information as a function of catalogue data features since network architectures (hidden and output size dimensions) could be kept fixed with the same initialized weights.

\end{document}